\pdfoutput=1
\documentclass[12pt,a4paper]{article}
\usepackage{jheppub}
\usepackage[usenames,dvipsnames]{xcolor}
 
\usepackage{microtype}
\usepackage{amssymb} 
\usepackage{float}
\usepackage{amsmath}
\usepackage{mathtools}
\usepackage{amsfonts}    
\usepackage{dsfont}
\usepackage{pdfpages}
\usepackage{verbatim}
\usepackage{esint}
\usepackage{array}

\usepackage{tensor}
\usepackage{mathrsfs}
\usepackage{textgreek} 
\usepackage[mathscr]{euscript}
\usepackage[smalltableaux]{ytableau}
\ytableausetup{centertableaux}
\usepackage{tikz}
\usetikzlibrary{decorations.pathreplacing, decorations.markings,calc,shapes.misc,decorations.pathmorphing,patterns.meta, math, shadings, backgrounds}
\usepackage{slashed}
\usepackage{datetime}
\usepackage{subcaption}

\definecolor{MathematicaBlue}{rgb}{0.368, 0.507, 0.710}
\definecolor{MathematicaYellow}{rgb}{0.881,0.611,0.142}
\definecolor{MathematicaGreen}{rgb}{0.560,0.692,0.195}
\definecolor{MathematicaRed}{rgb}{0.923,0.386,0.209}
\definecolor{MathematicaPurple}{rgb}{0.528,0.471,0.701}
\definecolor{light-gray}{gray}{0.75}

\usepackage{hyperref}
\hypersetup{
    pdfencoding=unicode,
	colorlinks=true,
	urlcolor=Maroon,
	linkcolor=RoyalBlue,
	citecolor=Maroon,
	pdftitle={Precision asymptotics of string amplitudes},
	pdfauthor={Marco M.\ Baccianti, Lorenz Eberhardt, Sebastian Mizera},
	pdfdisplaydoctitle=true,
	pdfstartview=FitH,
	linktocpage=true
}
\newcommand{\ba}{\begin{align}}

\newcommand{\be}{\begin{equation}}
\newcommand{\ee}{\end{equation}}
\def\bd{\begin{tikzpicture}}
\def\ed{\end{tikzpicture}}

\renewcommand\Im{\mathop{\text{Im}}}

\renewcommand\Re{\mathop{\text{Re}}}
\newcommand\Res{\mathop{\text{Res}}}
\newcommand\Disc{\mathop{\text{Disc}}}
\allowdisplaybreaks[1]
\DeclareMathAlphabet{\mathbbold}{U}{bbold}{m}{n}

\newcommand{\id}{\mathds{1}}

\newcommand\PSL{\text{PSL}}

\newcommand\SL{\text{SL}}

\newcommand{\M}{\mathcal{M}}

\newcommand\CC{\mathbb{C}}
\newcommand\ZZ{\mathbb{Z}}
\newcommand\RR{\mathbb{R}}
\newcommand\HH{\mathbb{H}}

\newcommand\QQ{\mathbb{Q}}

\renewcommand\d{\text{d}}
\newcommand\D{\mathrm{D}}

\newcommand{\Trop}{\mathrm{Trop}}
\renewcommand{\L}{\mathrm{L}}
\newcommand{\R}{\mathrm{R}}

\renewcommand{\le}{\leqslant}
\renewcommand{\ge}{\geqslant}
\renewcommand{\leq}{\leqslant}

\DeclareMathOperator*{\DRes}{DRes}

\title{Precision asymptotics of string amplitudes}

\author[1]{Marco Maria Baccianti}\emailAdd{m.m.baccianti@uva.nl}
\author[1]{\!\!, Lorenz Eberhardt}\emailAdd{l.eberhardt@uva.nl}
\author[2]{\!\!, Sebastian Mizera}\emailAdd{sebastian.mizera@columbia.edu}

\affiliation[1]{Institute for Theoretical Physics,
University of Amsterdam, Amsterdam, 1098XH, NL}
\affiliation[2]{Department of Physics, Columbia University, New York, NY 10027, USA}

\abstract{
Recent work revealed a tension between the Gross--Mende analysis of the high-energy fixed-angle behavior of string amplitudes and the explicit numerical data. Motivated by this puzzle, we revisit the problem of classifying saddle-point geometries for the one-loop amplitude.
We find an infinite family of complex saddles that dominate the high-energy regime.
Using general constraints and matching to numerical data, we formulate a bootstrap problem that determines their multiplicities. This procedure yields a precise asymptotic expansion of the one-loop amplitude at high energies. The resulting oscillatory contributions lead to a much richer high-energy behavior than that predicted by the original Gross--Mende analysis.
}

\begin{document}

\maketitle

\makeatletter
\g@addto@macro\bfseries{\boldmath}
\makeatother


\section{Introduction}

The study of string scattering amplitudes in the high-energy, fixed-angle limit has long served as a crucial window into the non-local nature of string theory. Unlike quantum field theories, where local interaction vertices lead to a polynomial behaviour of scattering amplitudes, the extended nature of the strings results in a qualitatively different, exponential suppression at fixed loop order. From the worldsheet perspective, this exponential damping is determined by specific saddle-point geometries, or equivalently, the classical solutions to the string equations of motion.

\paragraph{Gross--Mende analysis.}
In their seminal papers \cite{Gross:1987kza,Gross:1987ar}, Gross and Mende identified the most symmetric saddle point geometry at one-loop level, which looks like a tree-level saddle geometry ``folded'' on itself. As a result, the tension of the string effectively doubles, or equivalently, $\alpha'$ halves compared to tree level. The prediction for the $\alpha' s \to \infty$ asymptotics of the one-loop four-graviton amplitude $A(s,\theta)$ at fixed $\theta$ is thus
\be\label{eq:Gross--Mende prediction}
A(s,\theta) \sim \frac{1}{s^4 \sqrt{H(\theta)}} \mathrm{e}^{- \frac{1}{4}\alpha' s \mathcal{S}(\theta)}
\ee
compared to $\sim \mathrm{e}^{-\frac{1}{2}\alpha' s \mathcal{S}(\theta)}$ at tree level. Here, $\mathcal{S}(\theta)$ is a specific function of the scattering angle $\theta$,
\be
\mathcal{S}(\theta) = - \sin^2\big(\tfrac{\theta}{2}\big) \log \big(\sin^2(\tfrac{\theta}{2})\big)- \cos^2\big(\tfrac{\theta}{2}\big) \log \big(\cos^2(\tfrac{\theta}{2})\big)
\ee
obtained by evaluating the string action on the saddle. Similarly, the prefactor involves the determinant of the Hessian $H(\theta)$ which is an explicit expression.

Similar predictions have been extended to higher genus \cite{Gross:1987ar} and open strings \cite{Gross:1989ge}, as well as other external states \cite{Moeller:2005ez}. Gross--Mende analysis has been used to speculate on the resummation properties of string amplitudes \cite{Mende:1989wt}, which in turn has been qualitatively matched to the expectations from impact-parameter analysis of Amati--Ciafaloni--Veneziano \cite{Amati:1987wq,Amati:1987uf,Amati:1988tn} at the black hole/string correspondence point \cite{Atick:1988si,Susskind:1993ws, Horowitz:1996nw, Horowitz:1997jc, Damour:1999aw}. This transition has been studied from different points of view \cite{Veneziano:2004er,Giddings:2006vu, Giddings:2007bw}, see, e.g.\ \cite{Giddings:2011xs,DiVecchia:2023frv,Bedroya:2022twb} for reviews. More broadly, high-energy limit is important role in the S-matrix bootstrap \cite{Kruczenski:2022lot} and scattering amplitudes literature \cite{Alday:2007hr,Alday:2023pzu,Kervyn:2025wsb}. 

\begin{figure}
\centering
\hspace{-1.45em}\includegraphics[width=\textwidth]{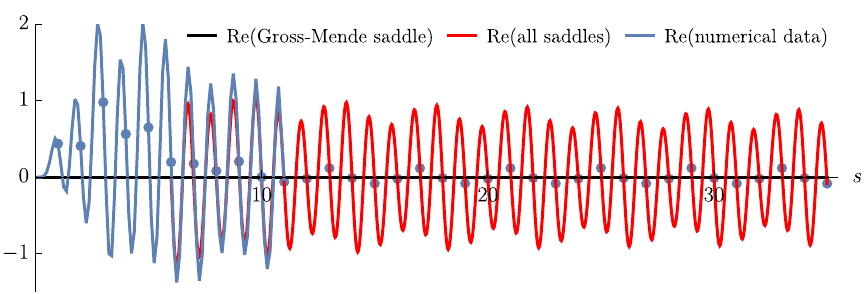}
\includegraphics[width=\textwidth]{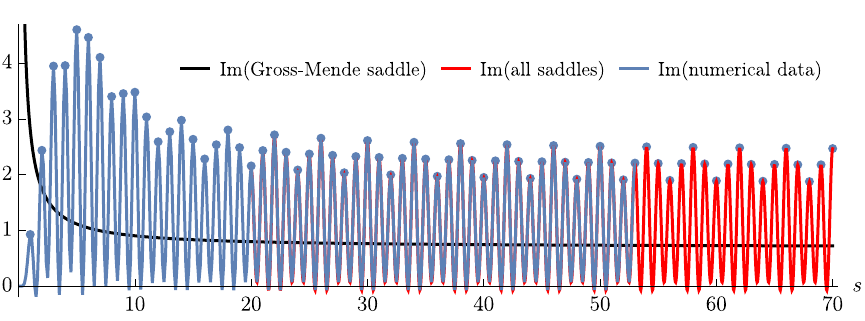}
\caption{\label{fig:intro}Comparison between numerical data (blue), the Gross--Mende saddle (black), and the sum over all saddles (red). The latter is obtained from the formula \eqref{eq:final formula} with $\alpha = 0$. The quantity plotted is the amplitude for fixed angle $\theta = \frac{\pi}{2}$ as a function of $s$, with stripped prefactors, $s^{4} \mathrm{e}^{s \mathcal{S}(\frac{\pi}{2})} \sin^2(\pi s) A(s,\frac{\pi}{2})$ where $\mathcal{S}(\tfrac{\pi}{2}) = \log(2)$ and $\alpha'=4$. The real and imaginary parts are plotted at the top and bottom respectively. Dots denote integer values of $s$ that can be more easily calculated due to a simplification of the Rademacher formula, as explained in section 4.3 of \cite{Baccianti:2025whd}.}
\end{figure}

\paragraph{Tension with data.}
Recently, in \cite{Baccianti:2025whd} we evaluated the amplitude $A(s,\theta)$ directly and found only a superficial agreement with \eqref{eq:Gross--Mende prediction} in the high-energy fixed-angle regime. The overall exponential envelope seemed to have been correct, but the details did not match. In particular, the Gross--Mende saddle predicts a purely imaginary amplitude, while in practice both the real and imaginary parts contribute at the same order of magnitude. This mismatch is illustrated in Fig.~\ref{fig:intro}. This prompted us to revisit Gross--Mende analysis.

\paragraph{What went wrong?}
Before diving into details, it is instructive to understand why the analysis of \cite{Gross:1987kza,Gross:1987ar} needs to be corrected. The path integral over all string geometries explicitly amounts to a complex integral over the moduli space of Riemann surfaces. Schematically, we can think of it as $\int \d^2 \tau\, f(\tau,\bar{\tau})$, where $\tau$ is the modular parameter specifying the shape of the genus-one Riemann surface, the torus. One cannot apply saddle-point analysis to such an integral directly, because it is not analytic and one cannot deform the integration contour to a steepest descent contour. Instead, one needs to first think of the integral as that over two independent variables $\tau$ and $\bar{\tau}$. Only then we are allowed to deform contours to pass through saddles. This, however, comes at a cost of a new complication when the analytic continuation $f(\tau,\bar{\tau})$ to independent $\tau$ and $\bar{\tau}$ has further complex saddles which do not lie on the original contour where $\bar{\tau}$ is the complex conjugate of $\tau$. This phenomenon opens a possibility of having new saddle point geometries.

In fact, the above issues are closely related to space-time signature and recent efforts to understand string scattering in the Lorentzian kinematics \cite{Eberhardt:2022zay,Eberhardt:2023xck,Eberhardt:2024twy,Banerjee:2024ibt,Baccianti:2025gll,Baccianti:2025whd}. The original Gross--Mende analysis was made in the unphysical kinematics where $\theta$ is purely imaginary. By contrast, we work with real $\theta \in (0,\pi)$ where one has to face the problem of extra complex saddles.

\paragraph{Methodology.}
The first observation is that there are infinitely many saddles with the same real part of the action $\mathcal{S}(\theta)$ as the saddle identified by Gross and Mende. In fact, there two classes of infinite families. The first one has to do with monodromies in the complexified modules parameter, $\tau$ and $\bar{\tau}$. Studying them will be the main point of this paper. The result is that saddle points are labelled by an element $\gamma$ of the congruence subgroup $\Gamma(2)$ of $\PSL(2,\ZZ)$ such that $(\tau,\bar{\tau})=(\gamma \cdot \tau_\theta,\tau_\theta^*)$ with $\tau_\theta$ the modular parameter of the Gross--Mende saddle. Here and in the rest of the paper, we reserve a star for actual complex conjugation, while barred quantities are not necessarily related by complex conjugation. For $\gamma \ne \id$, the corresponding saddles are complex. The second class has to do with monodromies in the puncture positions $z_i$, as they wind around each other. This winding appears already at tree level and is much simpler to analyze, see e.g.\ \cite{Yoda:2024pie}.

The next step is to determine the saddle point multiplicities, i.e., which saddles actually contribute and how many times. The responsible analysis would require understanding how to deform the complexified 8-dimensional integration contour into the paths of steepest descent originating at a set of saddles. It is known to be a formidable problem \cite{pham1983vanishing,arnold2012singularities}. To make progress, we instead set up a bootstrap problem in which we exploit symmetries of the problem to write an ansatz for multiplicities. We then combine it with numerical data we obtained using techniques developed in \cite{Eberhardt:2022zay,Eberhardt:2023xck,Banerjee:2024ibt,Baccianti:2025gll,Baccianti:2025whd}. This allows us to fix the ansatz almost entirely and make a prediction for the high-energy fixed-angle behavior of $A(s,\theta)$. 

\paragraph{Results.}
We claim that the asymptotics takes the form
\be\label{eq:introduction saddle formula}
A(s,\theta)\sim \sum_{\gamma} M_\gamma(s,\theta)\, A_\gamma(s,\theta) \ ,
\ee
where $A_\gamma(s,\theta)$ is the evaluation of the integral at the saddle labelled by $\gamma$ and $M_\gamma(s,\theta)$ is the corresponding multiplicity. The multiplicity itself can depend on the kinematics and it can be written as a polynomial in the phases $\mathrm{e}^{\pm \pi i s}$ and $\mathrm{e}^{\pm \pi i t}$, where $t = -s \sin^2(\tfrac{\theta}{2})$. We found that the sum over $\gamma$'s can be further split into two contributions: an infinite ``tail'' of saddles obtained by shifting the original Gross--Mende saddle in certain directions, as well as a finite number of exceptions.
These results are summarized in Tab.~\ref{tab:numerical multiplicities}, as well as plotted in Fig.~\ref{fig:intro}.
Under further favorable assumptions, we conjecture the final formula in \eqref{eq:final formula}, in which there is only one free multiplicity left that we haven't been able to determine conclusively.

Curiously, each individual term in \eqref{eq:introduction saddle formula} has bad analytic properties such as the lack of boundedness in the complex $s$-plane. However, we show that these are healed as a result of summing the infinite tail.

\paragraph{Outline.} This paper is organized as follows. In Section~\ref{sec:saddles} we give details of the saddle point evaluation and set up the problem of classifying saddles and their multiplicities. Before considering the full string theory amplitude, we analyze a toy model integral in Section~\ref{sec:toy integral} which illustrates which infinite families of saddles in $\tau$ we expect to find. We proceed with the string amplitude in Section~\ref{sec:saddle bootstrap} where we determine saddle multiplicities. We close with conclusions and outlook in Section~\ref{sec:conclusion}. In Appendix~\ref{app:toy integral} we give more details on the toy model integral from Section~\ref{sec:toy integral}. We also provide an extensive dataset of the amplitude on \texttt{Zenodo} \cite{Zenodo}, that we hope will also be useful for other studies.

\section{\label{sec:saddles}Saddles of the one-loop string amplitude}
The object of interest is the one-loop four-graviton  amplitude in type II string theory, which we call $A$. It takes the form
\be
A = \int_{\mathcal{F}} \frac{\d^2 \tau}{(\Im \tau)^5} \int_{\mathbb{T}^2} \prod_{j=1}^3 \d^2 z_j\ \prod_{1 \le j<i \le 4} |\vartheta_1(z_{ij} | \tau)|^{-2s_{ij}}\, \mathrm{e}^{\frac{2\pi s_{ij} (\Im z_{ij})^2}{\Im \tau}}\ . \label{eq:string integral}
\ee
We define $\d^2 x=\d \Re x\, \d \Im x$ for all integration variables. The Mandelstam invariants associated with the massless momenta $p_i$ are $s_{ij} = -(p_i + p_j)^2$ using mostly-plus signature. We use the conventions $\alpha'=4$. The integral proceeds over the fundamental domain $\mathcal{F}$ of the modular parameter $\tau$ and the torus $\mathbb{T}^2$ for the three punctures $z_i$, where $z_4 = 0$ is fixed by translation symmetry. The integrand involves the Jacobi theta function $\vartheta_1(z_{ij} | \tau)$ evaluated at $z_{ij} = z_i - z_j$. We stripped off trivial factors in \eqref{eq:string integral}, such as the polarization structure and the dependence on the string coupling.

To properly define the integral, one can employ the stringy $i \varepsilon$-prescription \cite{Witten:2013pra}. This prescription deforms the fundamental domain $\mathcal{F}$, now viewed as an the integration contour, near the boundaries of moduli space into a new contour $\mathcal{F}_{i\varepsilon}$. It is embedded in the complexified moduli space $\mathcal{M}_{1,4}^\CC$, where we take $z_i$ and $\tau$ to be independent from $\bar{z}_i$ and $\bar{\tau}$. As a consequence, the integrand has branch cuts on $\mathcal{M}_{1,4}^\CC$ and hence non-trivial monodromy properties.

Since the integrand is only invariant under modular transformations and lattice translations of the vertex operators acting jointly on left- and right-moving moduli, the complexified moduli space is globally defined as $\mathcal{M}_{1,4}^\CC=(\mathcal{S}_{1,4} \times \overline{\mathcal{S}}_{1,4})/\Gamma_{1,4}$. Here, 
\be 
\mathcal{S}_{1,4}\cong \{(\tau,z_1,z_2,z_3) \in \HH \times \CC^3  \, | \, z_i-z_j \not \in \ZZ \oplus \ZZ \tau\,,\  z_i \not \in \ZZ \oplus \ZZ \tau \}\ .
\ee 
is a covering of $\mathcal{M}_{1,4}$ on which all monodromies become abelian. It is parametrized by $\tau$ and the $z_i$'s. The group $\Gamma_{1,4} \cong \SL(2,\ZZ) \ltimes (\ZZ^2)^3$ is the group generated by modular transformations and shifts of $z_i$ along the lattice directions. The only remaining monodromy is given by braiding vertex operators around each other and is the same as for the tree-level integrand. Because of the braiding monodromies, the integrand is still a multi-valued function on $\mathcal{S}_{1,4}$ (for generic choices of kinematics), but this fact will not play a role for classification of the saddles, since each saddle in $z_i$ and $\tau$ will lead to an infinite family of saddles by applying braidings. These abelian phases can also be described using twisted homology \cite{Eberhardt:2024twy}.

We continue to write $\Im \tau=\frac{1}{2i}(\tau-\bar{\tau})$ in the following, which becomes a \emph{holomorphic} function on the complexification of moduli space. As already mentioned in the introduction, we reserve ${\bullet}^*$ for complex conjugation, while quantities with a bar $\bar{\bullet}$ are not necessarily the complex conjugate of the unbarred quantity.
The integration contour is then a half-dimensional contour in this complex moduli space, which one can think of describing complex metrics on the worldsheet. We refer to \cite{Baccianti:2025gll,Baccianti:2025whd} for further details of about this contour.

\subsection{Saddle point evaluation}
Gross and Mende \cite{Gross:1987kza} considered such amplitudes in a high energy limit, where $s,\, t \to \infty$ with the scattering angle $\theta$ given by $\sin^2(\frac{\theta}{2})=-\frac{t}{s}$ fixed. In this limit, the integrand sharply peaks around its saddle points and one can apply saddle point evaluation. Let us schematically write
\be 
A=\int_{\Gamma \subset \mathcal{M}_{1,4}^\CC} \d\mu( \boldsymbol{m})\, \mathrm{e}^{-s \, S(\boldsymbol{m},\theta)}\ ,  \label{eq:complex contour}
\ee
with $\boldsymbol{m}$ the collection of complexified moduli and $\d\mu(\boldsymbol{m})=\frac{\d^2 \tau}{(\Im \tau)^5} \prod_{j=1}^3 \d^2 z_j$. We refer to $S(\boldsymbol{m},\theta)$ as the action, since it can be identified with the on-shell action of the worldsheet theory. For saddles on the real locus $\mathcal{M}_{1,4}\subset \mathcal{M}_{1,4}^\CC$, this action is real by construction, but generically becomes complex away from the real locus.
To leading order in the saddle point approximation, one thus expects that the integral behaves like the integrand evaluated on the respective saddle, i.e.\ $\mathrm{e}^{-s S(\boldsymbol{m}_*,\theta)}$ if $\boldsymbol{m}_*$ denote the saddle moduli. Thus to get the leading asymptotics as $s \to \infty$, one needs to find the saddle $\boldsymbol{m}_*$ with the \emph{smallest} real part of the action. The smallest real part among all the saddle points turns out to be positive which leads to an exponentially decaying amplitude for physical kinematics.

However, there are in fact infinitely many saddles with the same real part of the saddle point action, but different imaginary parts of the saddle point action. Taking all these saddles into account will lead to an exponentially decaying amplitude overlaid with a delicate oscillatory structure. This leads to a far richer structure than what was originally anticipated by Gross and Mende, who identified one of these infinitely many saddles that lies on the real locus $\mathcal{M}_{1,4} \subset \mathcal{M}_{1,4}^\CC$. 

In principle, the path towards getting precise asymptotics for the string amplitude at high energies goes as follows: (i) find all the saddle points of $S(\boldsymbol{m},\theta)$ and (ii) deform the contour $\Gamma$ into a steepest descent contour and determine all the saddle multiplicities.
This is a \emph{very} hard task. As we shall see, there are infinitely many saddles and they do not follow a clear pattern. Determining which steepest descent contour through these saddles is homologically equivalent to the original contour $\Gamma$ seems hopeless to us, given the complicated topology of the moduli space.

Instead, we will largely bypass these steps and employ the following strategy. We will first classify all saddles with minimal $\Re S(\boldsymbol{m}_*,\theta)$. Even though there are also infinitely many of those, this turns out to be a doable problem. We will then apply various constraints, somewhat favorable assumptions and compare with numerical data obtained from evaluating the amplitude numerically at relatively large $s$ to bootstrap the saddle point multiplicities. This will lead us to the proposal for the behaviour of the amplitude given in \eqref{eq:introduction saddle formula}.

\subsection{Saddle hunting} \label{subsec:saddle hunting}
\begin{figure}
    \begin{minipage}{.325\textwidth}
        \includegraphics[width=\textwidth]{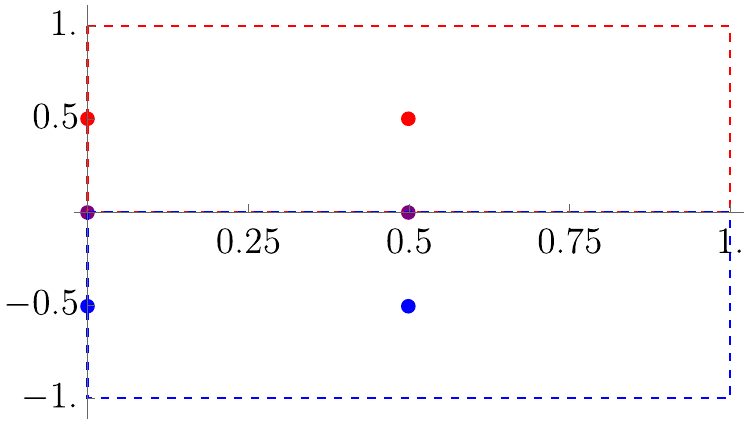}
    \end{minipage}
    \begin{minipage}{.325\textwidth}
        \includegraphics[width=\textwidth]{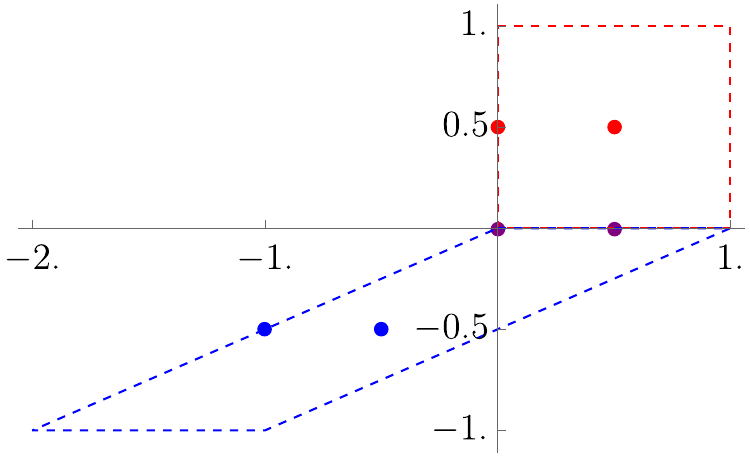}
    \end{minipage}
    \begin{minipage}{.325\textwidth}
        \includegraphics[width=\textwidth]{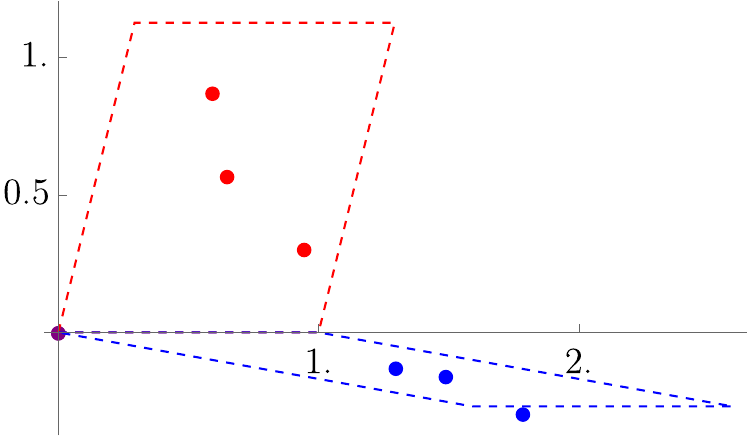}
    \end{minipage}
    \begin{minipage}{.325\textwidth}
        \includegraphics[width=\textwidth]{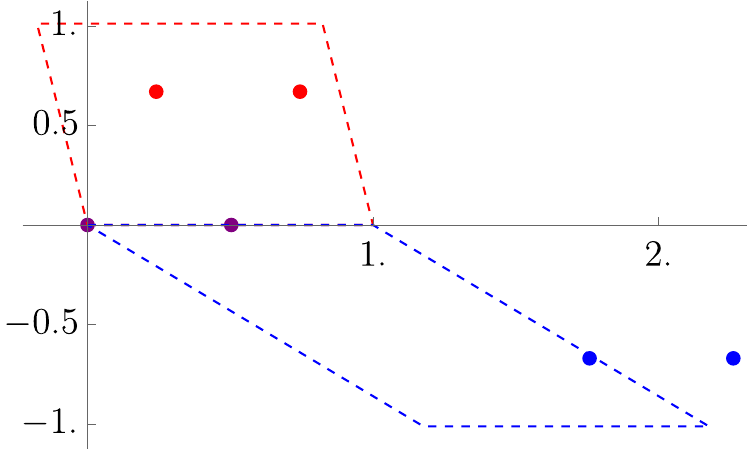}
    \end{minipage}
    \begin{minipage}{.325\textwidth}
        \includegraphics[width=\textwidth]{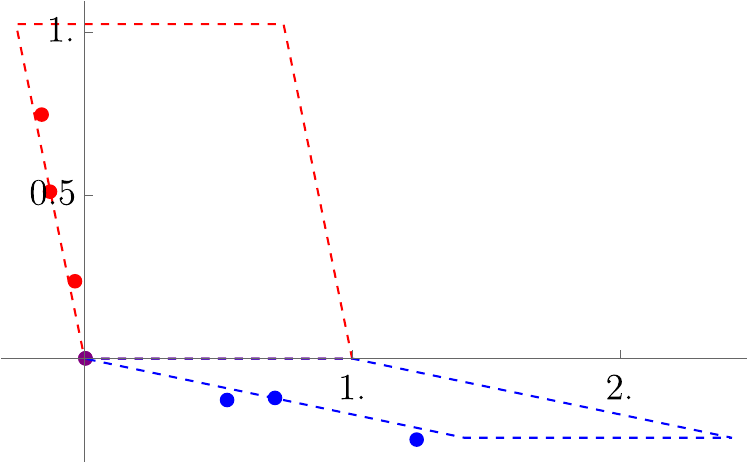}
    \end{minipage}
    \begin{minipage}{.325\textwidth}
        \includegraphics[width=\textwidth]{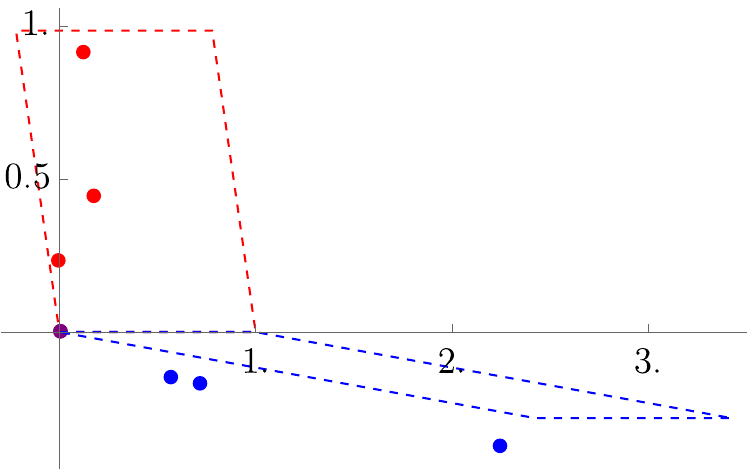}
    \end{minipage}
    \caption{A collection of sample saddles for $t=-\frac{s}{2}$. We draw the $z_i$'s as red points in the complex plane and the $\bar{z}_i$'s as blue dots in the complex plane. The red parallelogram is the fundamental domain of the torus described by $\tau$ (i.e.\ has vertices $0$, $1$, $\tau$ and $\tau+1$) and the blue parallelogram is the fundamental domain of the torus described by $\bar{\tau}$ (i.e.\ has vertices $0$, $1$, $\bar{\tau}$ and $\bar{\tau}+1$). The saddle point actions are in the order of display $\Re S(\boldsymbol{m}_*,\theta)\approx 0.693,\,  0.693, \, 0.996,\, 1.046,\,  1.109,\, 1.124$.
    The first saddle is the Gross--Mende saddle and is the only saddle we could find for which the right-moving moduli are the complex conjugates of the left-moving moduli. The second saddle has the same real of the action as the first saddle and is a complex version of the Gross--Mende saddle, where the locations of the vertex operators are still located at the half-periods of the parallelogram. The other four saddles have larger real part of the action, and are much more irregular. There are still some patterns, for example the punctures in the fourth saddle form a parallelogram, but this is not true for all saddles. The last saddle that we displayed doesn't seem to have any nice features at all.}
    \label{fig:sample saddles}
\end{figure}

The first part of our task is clear: we should identify the set of saddles of $S(\boldsymbol{m},\theta)$ on $\M_{1,4}^\CC$. The saddle point equations are not elementary, since they involve both elliptic functions as well as terms originating from the exponential in \eqref{eq:string integral}. This makes it impossible to give analytic solutions for the saddle points. Thus, our strategy is to perform a numerical search for saddles. Since there are infinitely many saddles, it is impossible to find them all in this way, but we identified about 2000 saddles numerically. A few sample saddles for $t=-\frac{s}{2}$ are displayed in Fig.~\ref{fig:sample saddles}. We have used the invariance under the modular group and lattice translations $\Gamma_{1,4}$ to move $\tau$ into the standard fundamental domain and $z_i$ for $i=1,2,3$ into the fundamental parallelogram with vertices $0$, $1$, $\tau$, $\tau+1$. We gauge fixed $z_4=\bar{z}_4=0$. 

Clearly, the set of all saddles is very complicated and a classification seems out of reach. Thus, we will restrict ourselves to the leading saddles for which $\Re S(\boldsymbol{m}_*,\theta)$ is as small as possible. For these, we observed in the data that all the vertex operators are inserted at half periods, i.e. $\{z_1,z_2,z_3,z_4\}$ is the set $\{0,\frac{1}{2},\frac{\tau}{2},\frac{\tau+1}{2}\}$ and $\{\bar{z}_1,\bar{z}_2,\bar{z}_3,\bar{z}_4\}$ is the set $\{0,\frac{1}{2},\frac{\bar\tau}{2},\frac{\bar\tau+1}{2}\}$. We can permute the half-periods by applying modular transformations. We can then choose for all our saddles that
\be 
(z_1, z_2, z_3, z_4) = \left( \frac{\tau+1}{2}\, , \frac{\tau}{2}\, , \frac{1}{2}\, , 0 \right)\, , \qquad (\bar{z}_1, \bar{z}_2, \bar{z}_3, \bar{z}_4) = \left( \frac{\bar{\tau}+1}{2}\, ,  \frac{\bar{\tau}}{2}\, ,  \frac{1}{2}\, ,  0 \right)\, .\label{eq:half period saddle point}
\ee
This permutation will be convenient when we discuss the contributing saddles in the $s$-channel.
We could not find a proof that all leading saddles are of this form, but our numerical search strongly suggest that this is the case.

The saddle point equations simplify dramatically if we insert this ansatz for the locations of the vertex operators. In fact, derivatives with respect to $z_i$ and $\bar{z}_i$ are zero. The saddle point equations for $\tau$ and $\bar{\tau}$ lead to the equation
\be\label{eq: saddle point equation gross mende lambda}
\lambda(\tau)=\lambda(-\bar{\tau})=-\frac{s}{t}\ ,
\ee
where
\be
\lambda(\tau)=\frac{\vartheta_2(\tau)^4}{\vartheta_3(\tau)^4}
\ee
is the modular lambda function.
In particular, it \emph{does not} imply that $\tau$ and $\bar{\tau}$ are related by complex conjugation. In fact, since $\lambda(\tau)$ is invariant under modular transformations of the principal congruence subgroup 
\be 
\Gamma(2)=\left\{\left. \begin{pmatrix}
    a & b \\ c & d
\end{pmatrix}\, \right|\, a \equiv d \equiv 1 \bmod 2\, , \ b \equiv c \equiv 0 \bmod 2 \right\}\ , \label{eq:Gamma(2) definition}
\ee
it follows that we can apply any modular transformation $\gamma \in \Gamma(2)$ acting only on left-moving coordinates and still obtain a saddle. We refer to the special solution where $\bar{\tau}$ is the complex conjugate of $\tau$ (i.e.\ $\bar{\tau}=\tau^*$) as the Gross--Mende saddle and to the others as complex Gross--Mende saddles.

\subsection{Complex Gross--Mende saddles} \label{subsec:complex Gross Mende saddles}
\paragraph{Covering surfaces.} The complex saddles that we identified above seem a bit exotic, so let us explain their geometric meaning. A torus admits a branched covering map to the sphere, branched over the half-periods of the torus. This covering map can be taken to be the Weierstrass $\wp$ function $\wp(z,\tau)$. As explained in \cite{Gross:1987kza}, this means that one can realize the Gross--Mende saddle as a double cover of the sphere, branched over the four points where the vertex operators are inserted.
\begin{figure}
    \centering
    \begin{tikzpicture}
            \draw[very thick] (0,-3) ellipse (1.5 and 1);
            \draw[very thick] (0,0) ellipse (1.5 and 1);
            \draw  (-1,-.4) node[cross out, draw=black, very thick, minimum size=5pt, inner sep=0pt, outer sep=0pt] {};
            \draw  (-1,-3.4) node[cross out, draw=black, very thick, minimum size=5pt, inner sep=0pt, outer sep=0pt] {};
            \draw  (.5,-.4) node[cross out, draw=black, very thick, minimum size=5pt, inner sep=0pt, outer sep=0pt] {};
            \draw  (.5,-3.4) node[cross out, draw=black, very thick, minimum size=5pt, inner sep=0pt, outer sep=0pt] {};
            \draw  (-.5,.4) node[cross out, draw=black, very thick, minimum size=5pt, inner sep=0pt, outer sep=0pt] {};
            \draw  (-.5,-2.6) node[cross out, draw=black, very thick, minimum size=5pt, inner sep=0pt, outer sep=0pt] {};
            \draw  (1,.4) node[cross out, draw=black, very thick, minimum size=5pt, inner sep=0pt, outer sep=0pt] {};
            \draw  (1,-2.6) node[cross out, draw=black, very thick, minimum size=5pt, inner sep=0pt, outer sep=0pt] {};
            \draw[very thick] (-1,-.4) to (-1,-3.4);
            \draw[very thick] (.5,-.4) to (.5,-3.4);
            \draw[very thick] (-.5,.4) to (-.5,-2.6);
            \draw[very thick] (1,.4) to (1,-2.6);
            \draw[very thick, decorate, decoration=snake] (-1,-.4) to (.5,-.4);
        		\draw[very thick, decorate, decoration=snake] (-.5,.4) to (1,.4);
            \draw[very thick, decorate, decoration=snake] (-1,-3.4) to (.5,-3.4);
        		\draw[very thick, decorate, decoration=snake] (-.5,-2.6) to (1,-2.6);
    \end{tikzpicture}
    \caption{The Gross--Mende saddle as a double cover.}
    \label{fig:Gross Mende double cover}
\end{figure}
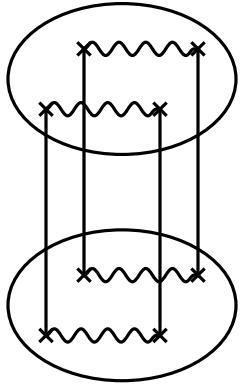
Schematically, this leads to the picture in Fig.~\ref{fig:Gross Mende double cover}. It is simple to see that this is a saddle as long as the tree-level configuration on the sphere was a saddle. In fact, it is useful to think about this saddle from a target space perspective. The embedding coordinate of the classical solution $X^\mu(z)$ into target space takes the same values on both sheets of the cover. Thus, the saddle configuration looks like the tree-level saddle that is superimposed on itself and the sheets are connected in a non-trivial way. The saddle surface is twice as small as the tree-level saddle because the it is proportional to the momentum on each sheet, which is half of the total momentum. From this argument, it is also simple to compute the on-shell action, which is proportional to the momentum squared on each sheet, but also integrated over all sheets. The former effect leads to a factor of $\frac{1}{4}$ with respect to the tree-level on-shell action, while the latter leads to a factor of $2$. Thus the on-shell action is half of the tree-level on-shell action, which translates in our conventions to
\be 
\mathcal{S}(\theta) := 
\Re S(\boldsymbol{m}_*,\theta) = - \sin^2\big(\tfrac{\theta}{2}\big) \log \big(\sin^2(\tfrac{\theta}{2})\big)- \cos^2\big(\tfrac{\theta}{2}\big) \log \big(\cos^2(\tfrac{\theta}{2})\big)>0\ . \label{eq:real part one loop action}
\ee
For example, $\mathcal{S}(\frac{\pi}{2}) = \log(2)$, matching with the numerical value $0.693$ mentioned below Figure~\ref{fig:sample saddles}.

\paragraph{Tree-level saddle point evaluation.} Via the double cover, the one-loop saddle is related to the tree-level saddle. Let us then discuss the tree-level saddle. It strictly speaking appears multiple times. The integrand is of the form $|z|^{-2s}|1-z|^{-2u}$ and we have a saddle at $z_*=\bar{z}_*=-\frac{s}{t}$. We can send
\be\label{eq:tree level braiding}
z \to \mathrm{e}^{2\pi i}z \qquad\text{or}\qquad 1-z \to \mathrm{e}^{2\pi i}(1-z)
\ee
(without the corresponding transformation for the right-movers $\bar{z}$) and produce a saddle that lives on a different sheet of the integrand when thought of as a holomorphic function in $z$ and $\bar{z}$ in the complexified moduli space $\mathcal{M}_{0,4}^\CC$. The saddles on the different sheets do contribute to the high-energy limit of the tree-level amplitude, which reads
\begin{align} 
A_\text{tree}(s,t)&=\frac{\Gamma(-s)\Gamma(-t)\Gamma(-u)}{\Gamma(1+s)\Gamma(1+t)\Gamma(1+u)}\\
&\sim -\frac{2 \sin(\pi t)\sin(\pi u)\, \mathrm{e}^{-2s\, \mathcal{S}(\theta)}}{\sin(\pi s) s\, t \, u}\ , \label{eq:tree-level asymptotics}
\end{align}
where in the second line we work in the $s$-channel regime $s > -t > 0$ and used the Stirling approximation of the Gamma function, together with the Euler reflection formula.

As promised, the amplitude falls off exponentially with the action appearing in the exponent that is twice as large as \eqref{eq:real part one loop action}. From a saddle point perspective, the trigonometric prefactors arise because one needs to consider saddles on different sheets, which appear with different phases. In fact, the $\sin(\pi s)$ in the denominator originates from a geometric series 
\be 
\frac{2i \sin(\pi t)\sin(\pi u)}{\sin(\pi s)}=1+\sum_{n=1}^\infty (\mathrm{e}^{-2\pi i t}+\mathrm{e}^{-2\pi i u}-2)\mathrm{e}^{2\pi i n s}\ . \label{eq:tree level trigonometric geometric series}
\ee
This infinite sum over different sheets appears because of the $i \varepsilon$-prescription that deforms the integration contour near $z=0$ into the Lorentzian direction where it traverses all the sheets labelled by the winding number $n$ of $z$ around the origin. 
\paragraph{Complex Gross--Mende saddles.} It now becomes clear why the complex Gross--Mende saddles discussed above are necessary. They appear once one complexifies the moduli and  passes to a different sheet in the base surface of the cover. In fact, we precisely have
\be 
\lambda(\tau+2)=\mathrm{e}^{2\pi i}\lambda(\tau)\ , \qquad 1-\lambda(\tfrac{\tau}{2\tau+1})=\mathrm{e}^{-2\pi i}\big(1-\lambda(\tau)\big)\ , \label{eq:lambda monodromies}
\ee
which parallel \eqref{eq:tree level braiding}.
The transformations $T^2:\tau \mapsto \tau+2$ and $\tilde{T}^2=ST^2S:\tau \mapsto \frac{\tau}{2\tau+1}$ are the two standard generators that generate $\Gamma(2)$ freely.
Contrary to the tree-level case, however, these different sheets do not merely modify the contribution of a given saddle by a phase. 

On top of this, we can also braid the locations of the vertex operators on the torus (while keeping $\tau$ and $\bar{\tau}$ fixed). This leads to the analogous phases to those that are already present at tree-level. We will deal with them after the fact by thinking of the saddle point multiplicities as combination of phases.

\subsection{Systematics of complex Gross--Mende saddles} \label{subsec:systematics}
Let us return to the Gross--Mende saddles in the $\tau$ and $z_i$ coordinates. We can define an inverse of $\lambda(\tau)$ through $\lambda^{-1}(z)=i\, \frac{{}_2 F_1(\frac{1}{2},\frac{1}{2},1;1-z)}{{}_2 F_1(\frac{1}{2},\frac{1}{2},1;z)}$. It maps a fundamental domain of $\Gamma(2)$ bijectively to $\CC \setminus \{0,1\}$. We define in particular
\be 
\tau_\theta=\lambda^{-1}(-\tfrac{s}{t})=i\, \frac{{}_2 F_1(\frac{1}{2},\frac{1}{2},1;1+\frac{s}{t})}{{}_2 F_1(\frac{1}{2},\frac{1}{2},1;-\frac{s}{t})}
= i\, \frac{{}_2 F_1(\frac{1}{2},\frac{1}{2},1,-\cot^2(\frac{\theta}{2}))}{{}_2 F_1(\frac{1}{2},\frac{1}{2},1,\csc^2(\frac{\theta}{2}))}\ . \label{eq: tau theta definition}
\ee
For standard branch cut choices of the hypergeometric function, this maps the interval $\theta \in (0,\pi)$ to the semicircle $\{|\tau+\frac{1}{2}|=\frac{1}{2}\} \cap \Im \tau> 0$ in the upper half plane.
For example, we have\footnote{For such special values, also the relevant modular functions that appear in the expressions below have closed-form evaluations in terms of special values of Gamma-functions, thanks to the Chowla--Selberg formula and the theory of complex multiplication. For example
\begin{subequations}
\begin{align} 
\vartheta_2(\tau_{\frac{\pi}{2}})&=\frac{\Gamma(\tfrac{1}{4})\mathrm{e}^{-\frac{\pi i}{8}}}{2^{\frac{1}{4}}\pi^{\frac{3}{4}}}\ , & \vartheta_3(\tau_{\frac{\pi}{2}})&=\frac{\Gamma(\tfrac{1}{4})\mathrm{e}^{-\frac{\pi i}{8}}}{2^{\frac{1}{2}}\pi^{\frac{3}{4}}}\ , & E_2(\tau_{\frac{\pi}{2}})&=\frac{6}{\pi}\ , \\
\vartheta_2(\tau_{\frac{5\pi}{6}})&=\frac{3^{\frac{1}{8}}\Gamma(\tfrac{1}{3})^{\frac{3}{2}}\mathrm{e}^{-\frac{\pi i}{12}}}{2^{\frac{1}{6}}\pi}\ , & \vartheta_3(\tau_{\frac{5\pi}{6}})&=\frac{3^{\frac{1}{8}} \Gamma(\tfrac{1}{3})^{\frac{3}{2}}\mathrm{e}^{-\frac{\pi i}{12}}}{2^{\frac{5}{12}}(\sqrt{3}-1)^{\frac{1}{2}}\pi}\ , & E_2(\tau_{\frac{5\pi}{6}})&=\frac{3 \Gamma(\frac{1}{3})^6\mathrm{e}^{\frac{2\pi i}{3}}}{2^{\frac{8}{3}}\pi^4}+\frac{4 \sqrt{3}}{\pi}\ .
\end{align} \label{eq:special values modular functions}%
\end{subequations}
}
\begin{subequations}
\begin{align}
t = -\frac{s}{2} \quad&\Leftrightarrow\quad \theta = \frac{\pi}{2} : \qquad \tau_{\frac{\pi}{2}} = \frac{-1+i}{2}\, ,\\
t=-\frac{(2+\sqrt{3})s}{4}\quad &\Leftrightarrow \quad 
\theta = \frac{5\pi}{6} : \qquad \tau_{\frac{5\pi}{6}} = \frac{-1+\sqrt{3}i}{4}\,.
\end{align}%
\end{subequations}
The $\tau_\theta$ from \eqref{eq: tau theta definition} will form our reference solution and any other solution can be obtained from it by applying a modular transformation in $\Gamma(2)$. By construction, another solution to $\lambda(\tau)=-\frac{s}{t}$ is given by $-\tau_\theta^*$, which is also related to $\tau_\theta$ by the following special transformation in $\Gamma(2)$,
\be 
-\tau^*_\theta=\frac{\tau_\theta}{2\tau_\theta+1}=\tilde{T}^2\cdot \tau_\theta\ .
\ee
We will fix in the following the freedom of performing joint modular transformations by putting $\bar{\tau}=\tau^*_\theta$ and $\tau=\gamma \cdot \tau_\theta$ for $\gamma \in \Gamma(2)$ and make $\gamma$ label different saddles.
There are a few more things to be noted.
\paragraph{Phases.} We have not discussed so far the imaginary part of the saddle point action. The on-shell action before inserting the value of $\tau$ and $\bar{\tau}$ reads
\begin{align} 
s S(\tau,\bar{\tau},\theta)&=s\log(\vartheta_2(\tau)^2)+t\log(\vartheta_3(\tau)^2)+u\log(\vartheta_4(\tau)^2)+(\tau \leftrightarrow -\bar{\tau}) \label{eq:on-shell action before tau saddle}\\
&=\frac{1}{2} \log \big(\lambda(\tau)^s(1-\lambda(\tau))^{u}\big)+(\tau \leftrightarrow -\bar{\tau})\ .
\end{align}
Thus, under $\tau \to \tau+2$, we pick up the phase $\mathrm{e}^{\pi i s}$, while under $\tau \to \frac{\tau}{2\tau+1}$, we pick up the phase $\mathrm{e}^{\pi i u}$. As discussed before, these phases are only understood to be up to multiplication by $\mathrm{e}^{2\pi i s}$ or $\mathrm{e}^{2\pi i t}$, which are produced by holomorphic braiding of the vertex operators. 
More generally, the phase of a saddle labelled by $\gamma \in \Gamma(2)$ is of the form
\be 
\mathrm{e}^{\pi i n s+\pi i m u}\ , \qquad n \in 2\ZZ+\frac{1}{2}\gamma_{1,2}\ , \qquad m \in 2\ZZ+\frac{1}{2}\gamma_{2,1}\ , \label{eq:phase constraint}
\ee
where $\gamma_{i,j}$ denotes the corresponding entry of the $2 \times 2$ matrix. The shift of $n$ arises from the fact that $\gamma_{1,2}$ defines the unique homomorphism $\Gamma(2) \longrightarrow \ZZ_2$ that gives $1$ for the generator $\tau \to \tau+2$ and $0$ for the generator $\tau \to \frac{\tau}{2\tau+1}$.
These phases lead to an oscillatory behaviour of the amplitude in the high energy limit exactly as for the tree level amplitude and thus agrees with the basic case explained above. A similar comment applies to the shift of $m$.

\paragraph{Crossing symmetry.} The crossing transformation $t \leftrightarrow u$, i.e.\ $\theta \leftrightarrow \pi-\theta$ preserves the $s$-channel. We have
\be 
\tau_{\pi-\theta}=-\frac{\tau_\theta+1}{2 \tau_\theta+1}=C \cdot \tau_\theta
\ee
with $C=(\begin{smallmatrix}
    -1 & -1 \\ 2 & 1
\end{smallmatrix}) \in \PSL(2,\ZZ)$. It satisfies $C^2=\id$. This means that under crossing transformations we get the new saddle 
\be 
(\tau,\bar{\tau})=(\gamma \cdot \tau_{\pi-\theta},\bar{\tau}_{\pi-\theta})=((\gamma C) \cdot \tau_\theta,C \cdot \bar{\tau}_\theta) \sim ((C \gamma C) \cdot \tau_\theta,\bar{\tau}_\theta)\ ,
\ee
where we used in the last step that saddles are defined up to modular transformations. In other words, the contribution from the saddle labeled by $C \gamma C$ is obtained from the saddle labeled by $\gamma$ by applying a crossing transformation. Provided that that there are no Stokes phenomena for $\theta \in (0,\pi)$, this means that the saddle multiplicities of such pairs of saddles are related by exchanging $t$ and $u$.

\paragraph{Reality.} Generically, the contribution from a given saddle is neither purely real nor purely imaginary. However, saddles also pair up as complex conjugate pairs. Under complex conjugation, left- and right-movers are exchanged and so the complex conjugate saddle is
\be 
(\tau,\bar{\tau})=(\tau_\theta,\gamma \cdot \tau^*_\theta)\sim (\gamma^{-1} \cdot \tau_\theta,\tau^*_\theta)\ .
\ee
Thus the saddle labelled by $\gamma^{-1}$ leads to the complex conjugate contribution.\footnote{One has to be careful about the choice of orientation of the contour through the saddle point. Depending on the choice of orientation, the contribution from the saddle labelled by $\gamma^{-1}$ is minus the complex conjugate of the contribution from the saddle $\gamma$.}

There is a further set of identifications on saddles that lead to complex conjugate contributions. In the above identification, we could have also not interchanged left- and right-movers, which gives the complex conjugate saddle
\be 
(\gamma \cdot \tau_\theta^*,\tau_\theta)=(\gamma \cdot (-(\tilde{T}^2 \cdot \tau_\theta)),\tilde{T}^{-2} \cdot (-\tau_\theta^*))\sim ((U \gamma U) \cdot \tau_\theta,\tau_\theta^*)\ ,
\ee
where $U=(\begin{smallmatrix}
    -1 & 0 \\ 0 & 1
\end{smallmatrix})\tilde{T}^{2}=(\begin{smallmatrix}
    -1 & 0 \\ 2 & 1
\end{smallmatrix}) \in \text{PGL}(2,\ZZ)$ (the group of integer matrices with determinant $\pm 1$, modulo $\pm \id$). The matrix $U$ satisfies $U^2=\id$ and thus also defines an outer automorphism on $\Gamma(2)$.

\paragraph{Automorphisms and multiplicities.} All complex Gross--Mende saddles have a $\ZZ_2$ automorphism originating from exchanging the two sheets of the covering. In terms of $\tau$ and $z_i$, the saddle is invariant under 
\be 
z_i \to -z_i\ , \qquad \bar{z}_i \to -\bar{z}_i
\ee
since for a half-period, we can translate the reflected puncture back to its original insertion point. In such a situation, since $\mathcal{M}_{1,4}^\CC$ is an orbifold as opposed to a manifold, saddle point multiplicities can be multiples of $\frac{1}{|\text{Aut}(\text{saddle})|}$, which is $\frac{1}{2}$ for all saddles. In order to not deal with half-integer multiplicities, we normalized the integral over $\mathcal{M}_{1,4}$ such that we get integer multiplicities. Indeed, one could multiply \eqref{eq:string integral} by an overall factor of $\frac{1}{2}$ accounting for the fact that $(\tau,z_1,z_2,z_3,z_4)$ and $(\tau,-z_1,-z_2,-z_3,-z_4)$ describe equivalent punctured tori. In the normalization of \eqref{eq:string integral}, all multiplicities will therefore be integer.

\subsection{Elementary saddles} \label{subsec:elementary saddles}
It will turn out that not all saddles appear in the final answer. In fact, conjecturally a relatively simple subset of saddles appears.
We call a saddle \emph{elementary} if we can write it has one of the following forms for $n \in \ZZ$,
\begin{subequations}
\begin{align}
    &(T^{2n}\cdot\tau_\theta,\tau^*_\theta)\ , \\
    &(T^{2n}\cdot\tau_\theta,-\tau_\theta)=(T^{2n}\cdot \tau_\theta,\tilde{T}^{2} \cdot \tau^*_\theta) \sim ((\tilde{T}^{-2} T^{2n}) \cdot \tau_\theta,\tau^*_\theta) \ , \\
    &(T^{2n}\cdot(-\tau^*_\theta),\tau^*_\theta)=((T^{2n}\tilde{T}^2)\cdot \tau_\theta,\tau^*_\theta)\ , \\
    &(T^{2n}\cdot(-\tau^*_\theta),-\tau_\theta)=((T^{2n}\tilde{T}^2)\cdot \tau_\theta,\tilde{T}^{2} \cdot \tau^*_\theta) \sim ((\tilde{T}^{-2} T^{2n}\tilde{T}^2) \cdot \tau_\theta,\tau^*_\theta)\ ,
\end{align} \label{eq:elementary saddle list}%
\end{subequations}
i.e.\ $\gamma$ is of the form $T^{2n}$, $\tilde{T}^{-2}T^{2n}$, $T^{2n}\tilde{T}^2$ or $\tilde{T}^{-2}T^{2n}\tilde{T}^2$ for $n \in \ZZ$. Recall that $T^2 : \tau \mapsto \tau + 2$ and $\tilde{T}^2 : \tau \mapsto \frac{\tau}{2\tau + 1}$ are the two generators of $\Gamma(2)$.
This set of saddles closes under complex conjugation, $\gamma \mapsto \gamma^{-1}$ and $\gamma \mapsto U \gamma U$, and crossing. For the latter two properties, one observes that $U T^{2n}U=\tilde{T}^{-2}T^{-2n}\tilde{T}^2$, $U \tilde{T}^{-2}T^{2n}U=T^{-2n}\tilde{T}^{-2}$, $C T^{2n} C=\tilde{T}^{-2}T^{2n}\tilde{T}^2$ and $C \tilde{T}^{-2}T^{2n}C=T^{2n-2}\tilde{T}^{2}$.
Taken together, this implies also that the saddles $T^{2n}$ and $\tilde{T}^{-2}T^{2n}\tilde{T}^2$ give actually identical contributions which are automatically crossing symmetric. We will thus omit the latter in the following.

We now give two heuristic reasons why only this set of saddles ultimately contributes to the string amplitude. 

\paragraph{Approximate holomorphic factorization.} Let us return to the original integral representation \eqref{eq:string integral} of the string amplitude. The zero mode term of the Green's functions $\exp \sum_{1 \le j<i \le 4}\frac{2\pi s_{ij} (\Im z_{ij})^2}{\Im \tau}$  prevents the integrand from holomorphically factorizing. If we momentarily ignore the presence of this term, then the integrand would be invariant up to phases under $\tau \to \tau+2$, $z_1 \to z_1+1$, $z_2 \to z_2+1$ (without applying this shift to the right-moving coordinates). The transformation on the vertex operator positions is dictated by the relation \eqref{eq:half period saddle point} under the shift. The steepest descent contour through each of these saddle points would thus be identical. One could then first perform the integral over the vertex operator positions $z_i$ via saddle point approximation, which would then effectively impose the relations \eqref{eq:half period saddle point}. The remaining integral over $\tau$ and $\bar{\tau}$ would then to leading order be given by \eqref{eq:on-shell action before tau saddle}
\be 
|\vartheta_2(\tau)|^{-4s}|\vartheta_3(\tau)|^{-4t}|\vartheta_4(\tau)|^{-4u}+\text{perms of $(s,t,u)$}\ . \label{eq:toy integrand main text}
\ee
We will consider an integral of this quantity later on in Section~\ref{sec:toy integral} as a toy integral for the full amplitude. In this case, the steepest descent contour can be determined exactly. It will allow us to see why only the elementary saddles \eqref{eq:elementary saddle list} contribute.

The full string amplitude integrand, however, \emph{does not} holomorphically factorize which makes the identification of relevant saddles more complicated. Nevertheless, the argument can be partially rescued since the integral \emph{approximately} factorizes in the following sense. Let us put $z_1=\frac{\tau+1}{2}+\delta z_1$, $z_2=\frac{\tau}{2}+\delta z_2$ with $\delta z_1$ and $\delta z_2$ of order 1. For the steepest descent contour near the elementary saddles, $\delta z_1$, $\delta z_2$, $z_3$, $z_4$, $\bar{z}_i$ and $\bar{\tau}$ to be of order one and $\tau$ is large for large $|n|$ (where $n$ is as in \eqref{eq:elementary saddle list}). We also gauge fix $z_4=\bar{z}_4=0$. Then the argument of the exponential becomes
\be 
\sum_{1 \le j<i \le 4}\frac{2\pi s_{ij} (\Im z_{ij})^2}{\Im \tau} =\frac{\pi i s (\bar{\tau}-\tau)}{2}+\pi i s (z_{13}+z_{24}-\bar{z}_{13}-\bar{z}_{24})+\mathcal{O}(\tau^{-1})\ .
\ee
The large and order 1 terms \emph{do} holomorphically factorize. This means that the previous argument is correct for large enough $n$. Thus, the steepest descent contour will stabilize for large $n$, which means that only elementary saddles can contribute, except for possibly a finite number of non-elementary saddles for which the distance of $\gamma\cdot \tau$ and $\bar{\tau}$ is small and for which this argument fails. The bootstrap that we consider below strongly suggests that no such exceptions exist. We will recycle this argument below since it also implies that the saddle point multiplicities have to stabilize for large enough $n$. 

\paragraph{Inherited tree-level saddles.} One can also motivate the appearance of elementary saddles by comparing with the tree-level saddle via the double cover realization discussed in Section\ref{subsec:complex Gross Mende saddles}. For tree-level saddles, also only a small subset of saddles contributed to the final answer which one can read off from \eqref{eq:tree level trigonometric geometric series}. The only appearing phases are of the form $\mathrm{e}^{2\pi i n s+2\pi i m u}$ with $n \in \ZZ$ and $m \in \{-1,0,1\}$ (in fact the range of $n$ is also restricted, which we will come back to later). Via the monodromies of $\lambda$ \eqref{eq:lambda monodromies}, we see that these saddles precisely uplift to the elementary saddles in the double covering surface.\footnote{Since the tree-level integrand has abelian monodromies, one cannot distinguish $T^{2n}$ from $\tilde{T}^{-2}T^{2n}\tilde{T}^2$ by looking at the phases alone. Both appear as one can see by considering the actual steepest descent contour.} In fact, the multiplicities that we will find for the one-loop amplitude take a very similar form to \eqref{eq:tree level trigonometric geometric series}, but the series will not be a geometric series. Thus, it is essentially the nicest possible scenario if the saddles contributing to the one-loop string amplitude are directly inherited from the tree-level saddles.

\subsection{Evaluating the contribution from a single saddle}
After having obtained a good grasp over the appearing saddles, we now want to evaluate the contribution from a single saddle point, i.e.\ an integral of the form
\be 
\int \d^8 \boldsymbol{m} \, f(\boldsymbol{m})\, \mathrm{e}^{-s S(\boldsymbol{m},\theta)}
\ee
with $f(\boldsymbol{m})=(\Im \tau)^{-5}$ and
\be 
s S(\boldsymbol{m},\theta)=\sum_{1 \le j<i \le 4} s_{ij} \bigg(\log |\vartheta_1(z_{ij}|\tau)|^2-\frac{2\pi (\Im z_{ij})^2}{\Im \tau}\bigg)\ .
\ee
It will in the following also be convenient to write $\tilde{\tau}=-\bar{\tau}$.
We want to compute this integral as an asymptotic expansion in $s^{-1}$ around a saddle point. As mentioned above, we will consider the saddle point \eqref{eq:half period saddle point} and $\lambda(\tau)=\lambda(\tilde{\tau})=-\frac{s}{t}$. We will express the following expressions as functions of $\tau$ and $\tilde{\tau}$. The saddle point evaluation can be done by the usual Feynman diagram rules, where we view $f(\boldsymbol{m})$ as an operator insertion.
To first subleading order in the $s^{-1}$ expansion expansion, we obtain
\begin{multline}
    \pm \frac{(2\pi)^{n/2}f \mathrm{e}^{-s S}}{s^{n/2}\sqrt{\det H_{ij}}} \Bigg(1+s^{-1}\bigg[ H^{ij} H^{k\ell}H^{pq} \Big(\frac{1}{8} A_{ijk}A_{lpq}+\frac{1}{12} A_{ikp}A_{jlq}\Big)-\frac{1}{8} H^{ij} H^{kl} K_{ijkl}\\
   -\frac{1}{2} H^{ij} H^{kl} f^{-1}\partial_l f A_{ijk}+\frac{1}{2} H^{ij} f^{-1} \partial_i \partial_j f\bigg]+\mathcal{O}(s^{-2})\Bigg)\ , \label{eq:general saddle point correction}
\end{multline}
where $H_{ij}=\partial_i \partial_j S$, $H_{ij} H^{jk}=\delta_i^k$, $A_{ijk}=\partial_i \partial_j \partial_k S$, $K_{ijkl}=\partial_i \partial_j \partial_k \partial_l S$ with $i,j,k,l=1,\ldots,8$ and all quantities are assumed to be evaluated on the saddle point, where the fractional prefactors are symmetry factors of the respective diagram. The overall $\pm$ sign can be thought of as the freedom of defining the square root of $\det H_{ij}$ or of the choice of orientation through the saddle point. It is useful to also evaluate the subleading correction for the bootstrap below, since it will allow us to use numerical data for lower values of $s$. We will denote the contribution of a particular saddle point labelled by $\gamma \in \Gamma(2)$ with $A_\gamma(s,\theta)$.

\paragraph{Theta identities.} These expressions require us to compute a large number of derivatives of Jacobi theta functions, both in $z_i$ and $\tau$ (as well as the right-moving coordinates). Derivatives of modular forms are in general not modular, but only quasimodular. In fact, let us recall that the second Eisenstein series
\be 
E_2(\tau)=1-24 \sum_{n=1}^\infty \frac{nq^n}{1-q^n}\ , \qquad q=\mathrm{e}^{2\pi i \tau}
\ee
is a quasi-modular form satisfying
\be 
E_2(\tfrac{a \tau+b}{c \tau+d})=(c \tau+d)^2 E_2(\tau)-\frac{6 i c(c \tau+d)}{\pi}\ .
\ee
We also recall that $E_2(\tau)$ admits a non-holomorphic modular completion of the form
\be 
E_2^*(\tau)=E_2(\tau)-\frac{3}{\pi \Im \tau }\, ,
\ee
which is a non-holomorphic modular form of weight $(2,0)$. In fact, we have
\be 
\partial_\tau \vartheta_2(\tau)=\frac{\pi i}{12}\vartheta_2(\tau) \big(E_2(\tau)+\vartheta_3(\tau)^4+\vartheta_4(\tau)^4\big)
\ee
with similar formulas for the derivative of $\vartheta_3$ and $\vartheta_4$. This means that we can express the result of the saddle point approximation in terms of Jacobi theta functions and the second Eisenstein series and $\Im \tau$. Since the final result is modular, the second Eisenstein series only appears in its non-holomorphic version $E_2^*(\tau)$.

\paragraph{Hessian.} By direct computation one finds for the determinant of the Hessian
\be 
\det H=2^8\pi^{10}\left|\frac{\vartheta_3(\tau)^8\vartheta_4(\tau)^8}{\vartheta_2(\tau)^8}\right|^2 \prod_{i=1}^3 \left(\frac{\pi^2 }{9} |E_2^*(\tau)+F_i(\tau)|^2-\frac{1}{(\Im \tau)^2}\right) \label{eq:determinant Hessian}
\ee
with
\be 
F(\tau)=\begin{pmatrix} -\vartheta_2(\tau)^4-\vartheta_3(\tau)^4\\
\vartheta_2(\tau)^4-\vartheta_4(\tau)^4 \\
\vartheta_3(\tau)^4+\vartheta_4(\tau)^4
\end{pmatrix}\ .
\ee
$F(\tau)$ transforms as a vector-valued modular form of weight 2, where the representation matrix permutes the three elements. We also use the notation $| \bullet|^2=\bullet \bar{\bullet}$, where $\bar{\bullet}$ denotes the corresponding right-moving quantity where one exchanges $\tau$ and $\tilde{\tau}$. Thus, the Hessian written in this form is modular covariant. The non-invariance of the theta function prefactor is compensated by the covariance of $f(\boldsymbol{m})=(\Im \tau)^{-5}$ and the fact that modular transformations also permute non-trivially $(s,t,u)$. This form is remarkably compact given that we computed the determinant of a relatively complicated $8 \times 8$ matrix. To leading order, we thus have
\be 
A_\gamma(s,\theta)=\frac{(2\pi)^4i\, \mathrm{e}^{-s \, \mathcal{S}(\theta)}}{s^4(\Im \tau)^5 \sqrt{-\det H}}\bigg|_{\tau=\gamma \cdot \tau_\theta,\, \bar{\tau}=\tau^*_\theta}\ , \label{eq:Agamma expression}
\ee
where $\mathcal{S}(\theta)$ is given in \eqref{eq:real part one loop action}. These contributions are accompanied by phases which we prefer to keep separate and which are constrained according to \eqref{eq:phase constraint}.
For the Gross--Mende saddle, the determinant of the Hessian turns out to be negative. Thus, we included an extra minus sign in the square root and compensated by an $i$ in order to make the sign choice canonical. In general, we will choose the principal branch in \eqref{eq:Agamma expression} for $\theta=\frac{\pi}{2}$ and then smoothly follow the branch as a function of $\theta$. 

\paragraph{Subleading order correction.} We also worked out the explicit form of the $\mathcal{O}(s^{-1})$ correction from \eqref{eq:general saddle point correction}. It simplifies somewhat and can be written as a polynomial of order 8 in $E_2^*(\tau)$ and $E_2^*(\tilde{\tau})$. One of the a priori three factors of $\det H$ in the denominator originating from the three inverse matrices $H^{ij}$ in \eqref{eq:general saddle point correction} cancels. The precise formula can be found in a \texttt{Mathematica} notebook on \texttt{Zenodo} \cite{Zenodo}.

\section{Toy integral} \label{sec:toy integral}

In this section, we explain some general features of the saddle point approximation with the help of a toy integral. This toy integral is obtained by performing the saddle point approximation of the $z_i$ integrals. The full string integral exhibits many of the features of this toy integral because of the approximate holomorphic factorization that we explained in Section~\ref{subsec:elementary saddles}. 

\subsection{On-shell action}
We take as an integrand of the toy integral the on-shell action \eqref{eq:on-shell action before tau saddle} before we take the saddle in $\tau$, 
\begin{align}
    \mathcal{A}^{\text{toy}}(s,
    \theta)&=\int_{\mathcal{F}_{i\varepsilon}} \frac{\d^2 \tau}{\Im(\tau)^2} \big(|\vartheta_2(\tau)|^{-4s}|\vartheta_3(\tau)|^{-4t}|\vartheta_4(\tau)|^{-4u}+\text{perms of $(s,t,u)$}\big) \label{eq: definition of toy amplitude}\\
    &=\int_{\mathcal{F}_{i \varepsilon}^{\Gamma(2)}} \frac{\d^2 \tau}{\Im(\tau)^2} \, |\vartheta_2(\tau)|^{-4s}|\vartheta_3(\tau)|^{-4t}|\vartheta_4(\tau)|^{-4u}\ . \label{eq:toy integral}
\end{align}
In the second line, we use that we can rewrite the sum over the 6 permutations of $(s,t,u)$ as an integral over the fundamental domain $\mathcal{F}^{\Gamma(2)}$ of the congruence subgroup $\Gamma(2)$ defined in \eqref{eq:Gamma(2) definition}, with a suitable $i \varepsilon$ deformation near the cusp.
The strategy to evaluate \eqref{eq: definition of toy amplitude} is to holomorphically split the contour integral over $\tau$ and $\bar{\tau}$. We will in the following write $\tilde{\tau}=-\bar{\tau}$ so that both $\tau$ and $\tilde{\tau}$ are in the upper half plane. It was shown in \cite{Baccianti:2025gll} by using that the fundamental domain of $\Gamma(2)$ can be mapped to a cross ratio via the modular lambda function $\lambda$. In fact, when following the same manipulations as in \cite[Section 2.2]{Baccianti:2025gll}, one can split the contour as follows (assuming that $s$, $t$ and $u$ describe physical kinematics),\footnote{In \cite{Baccianti:2025gll}, the opposite sign of the $i \varepsilon$ prescription was employed since only the real part was of interest. As a consequence, the $\tau$-contour goes to $-\infty+i$ in our case instead of to $\infty+i$. For $s$-channel kinematics, the integrand is non-singular as $\tau$ or $\tilde{\tau}$ approach on of the other cusps of $\mathcal{F}^{\Gamma(2)}$ located at $0$ and $1$. Thus contrary to the most general case discussed in \cite{Baccianti:2025gll}, the direction from which these cusps are approached is not important. }
\be 
 \mathcal{A}^{\text{toy}}(s,\theta) =  \frac{1}{2i}\int_0^{-\infty+i} \d \tau  \int_{-1}^1 \d \tilde{\tau}\, \frac{ |\vartheta_2(\tau)|^{-4s}|\vartheta_3(\tau)|^{-4t}|\vartheta_4(\tau)|^{-4u}}{\Im(\tau)^2}  \ .\label{amplitude toy with hol split contour}
\ee
Here, the contours connect the indicated endpoints in the upper half plane. 
 Since the integrand holomorphically factorizes for large $s$, we can analyze the steepest descent contour separately in $\tau$ and $\bar{\tau}$.
The imaginary part can be evaluated more efficiently by computing the difference between the two $i\varepsilon$ prescriptions, which leads to:
\begin{align}\label{eq: toy integral im part formula}
    \Im \mathcal{A}^{\text{toy}}(s,\theta) = \frac{1}{4} \int_{-\infty + i}^{\infty + i} \d \tau \int_{-1}^{1} \d \tilde{\tau} \,  \frac{ |\vartheta_2(\tau)|^{-4s}|\vartheta_3(\tau)|^{-4t}|\vartheta_4(\tau)|^{-4u}}{\Im(\tau)^2} \ .
\end{align}

\subsection{Steepest descent contour}
\begin{figure}
\centering
        \includegraphics[width=\linewidth]{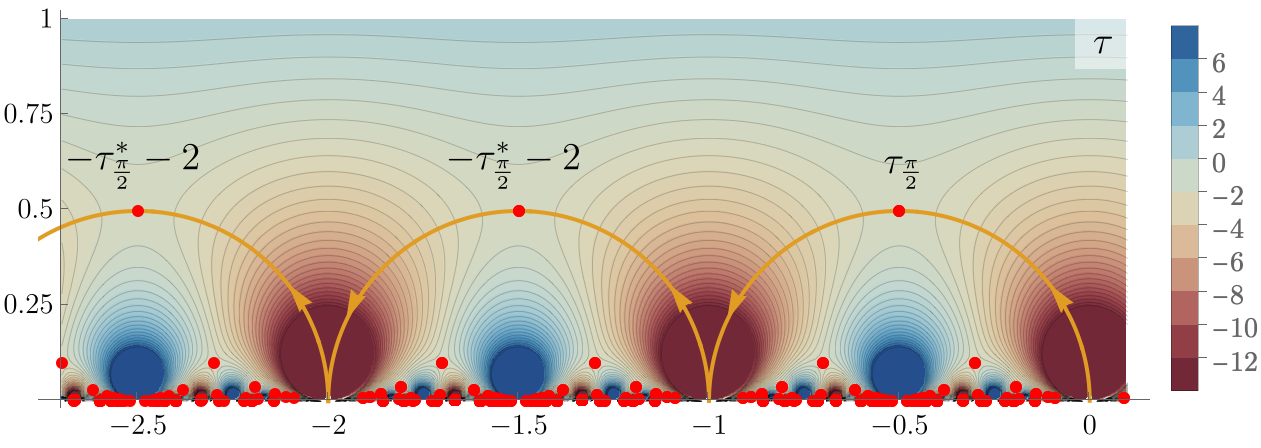}
        \includegraphics[width=\linewidth]{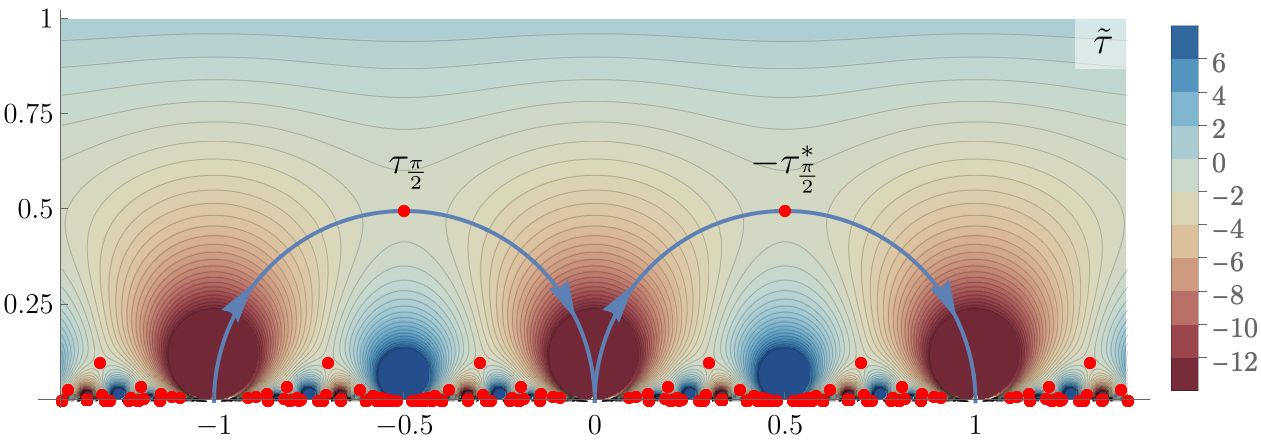}
     \caption{The contour employed in \eqref{amplitude toy with hol split contour} when deformed to the steepest descent contour in the $\tau$ (top, orange) and $\tilde\tau$ (bottom, blue) upper-half planes. 
    The background shows level curves of the absolute value of the integrand for $s=5$ and $\theta=\frac{\pi}{2}$ on a logarithmic scale. Saddles are marked with red dots.}
    \label{fig: re part toy all}
\end{figure}

Now that we know what the integration contours are, the next step is to understand where the saddle points lie in $\HH\times\HH$, and which of these contribute. In fact, the saddle points for our toy integral are by construction identical to the saddle points of the full string amplitude which we already discussed in Section~\ref{subsec:systematics}. Thus, up to modular transformations, they are given by $\tilde{\tau}=-\tau_\theta^*$ and $\tau=\gamma \cdot \tau_\theta^*$ for $\gamma \in \Gamma(2)$, with the same identifications as in Section~\ref{subsec:systematics}.
To determine which saddle points contribute with what multiplicity, we have to determine the steepest descent contour, which can be done independently in $\tau$ and $\tilde{\tau}$, since the integrand factorizes holomorphically. The absolute value of $|\vartheta_2(\tau)^{-2s}\vartheta_3(\tau)^{-2t}\vartheta_4(\tau)^{-2u}|$ is drawn in Figure~\ref{fig: re part toy all}.
Thus we see that the saddle points in $(\tau,\bar{\tau})$ that contribute are elementary as discussed in \ref{subsec:elementary saddles}, see Equation~\eqref{eq:elementary saddle list}. To each elementary saddle there is an associated modular transformation $\gamma$ and a phase $\mathrm{e}^{\pi i n s+\pi i mu}$ that the integrand will pick, as described in \ref{subsec:systematics}. 
Explicitly the relevant $(\tau_{\text{saddle}},\bar{\tau}_{\text{saddle}},\mathrm{e}^{\pi i n s+\pi i mu})$, with $n \in \ZZ_{\ge 0}$ are:
\begin{subequations}
\begin{align} 
\gamma&=T^{-2n}\ , & &(\tau_\theta-2n,\tau_\theta^*;\mathrm{e}^{\pi i n s})\ , \\
\gamma&=\tilde{T}^{-2}T^{-2n}\ , & & (\tau_\theta-2n,-\tau_\theta;\mathrm{e}^{\pi i n s-\pi i u})\ , \\
\gamma&=T^{-2n-2}\tilde{T}^2\ , & &(-\tau_\theta^*-2n-2;\tau_\theta^*;\mathrm{e}^{\pi i n s-\pi i t})\ , \\
\gamma&=\tilde{T}^{-2}T^{-2n-2}\tilde{T}^2\ , & &(-\tau_\theta^*-2n-2;-\tau_\theta;\mathrm{e}^{\pi i (n+1) s})\ ,
\end{align}\label{eq:elementary saddles toy integral}%
\end{subequations}
and they all contribute with unit multiplicity.

The phases can be easily determined by noting that under $\tau \to \tau+2$, the integrand $\vartheta_2(\tau)^{-2s}\vartheta_3(\tau)^{-2t}\vartheta_4(\tau)^{-2u}$ will pick up the phase $\mathrm{e}^{-\pi i s}$ coming from the leading term in the $q$-expansion of $\vartheta_2$.

Similarly, the phase under the transformation $\tau \rightarrow \frac{\tau}{2 \tau+1}$ can be obtained via the identity $\vartheta_2(\tau)^{-2s}\vartheta_3(\tau)^{-2t}\vartheta_4(\tau)^{-2u}=\lambda(\tau)^{-\frac{s}{2}}\left(1-\lambda(\tau) \right)^{-\frac{u}{2}}$ and by looking at the leading term of the $q$-expansion of $\lambda(-\frac{1}{\tau})=1-\lambda(\tau)$.

\subsection{Saddle point evaluation}
We can now directly evaluate the toy integral \eqref{amplitude toy with hol split contour} in the large $s$ limit.

\begin{figure}
        \centering
\raisebox{.4cm}[\height][0pt]{\includegraphics[width=.49\linewidth]{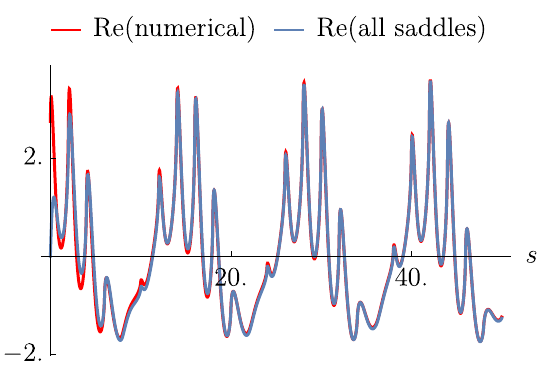}}
\includegraphics[width=.49\linewidth]{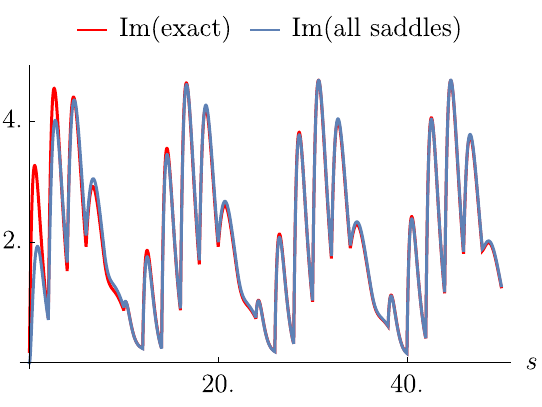}
    \caption{Comparison of the direct evaluation of the toy integral \eqref{eq: definition of toy amplitude} (in red) vs the saddle point expression (in blue). We chose the angle $\theta=\frac{\pi}{4}$ in these plots. The plot on the left is the real part, and on the right the imaginary part. The imaginary part can be computed exactly as described in Appendix~\ref{app: im part exact toy}, while the numerical curve for the real part was generated as described in Appendix~\ref{subapp:numerical evaluation}. The relevant mathematica notebook is also attached to the \texttt{Zenodo} submission \cite{Zenodo}.
    }
    \label{fig: plots toy}
\end{figure}

\paragraph{Contribution of a single saddle.} The Hessian of a single saddle in the variables $(\tau,\bar{\tau})$ is much simpler than for the full string amplitude because the integrand (excluding the $s$-independent term $(\Im \tau)^{-2}$ from the measure) holomorphically factorizes. Thus the determinant of the Hessian factorizes as follows,
\be 
\det H^{\text{toy}}(\tau,\tilde{\tau})=\Big|\frac{1}{s}\partial_\tau^2\log \big(\vartheta_2(\tau)^{-2s}\vartheta_3(\tau)^{-2t}\vartheta_4(\tau)^{-2u}\big)\Big|^2 =\frac{\pi^4}{4}\cot\left( \frac{\theta}{2}\right)^4 |\vartheta_3(\tau)^8|^2 \ .
\ee
We then get the following contribution from a single saddle, 
\begin{align}
    A^{\text{toy}}_{\gamma}(s,\theta) =\frac{i\, \mathrm{e}^{-s \mathcal{S}(\theta)}\tan(\frac{\theta}{2})^2}{s\pi (\Im \tau)^2  |\vartheta_3(\tau)^4|^2}\bigg|_{\tau=\gamma \cdot \tau_\theta,\, \bar{\tau}=\tau^*_\theta}\ , \label{eq:Agamma toy expression}
\end{align}
with $\mathcal{S}(\theta)$ given in \eqref{eq:real part one loop action}. This expression does not yet include the phase prefactor that we view as part of the multiplicity and is specified in \eqref{eq:elementary saddle list}.  \paragraph{Sum over saddles.} When summing over the saddles determined above, the saddle point approximation reads
\begin{align}\label{eq: im toy saddle 3 terms}
     \mathcal{A}^{\text{toy}} \sim  -A_{\id}^{\text{toy}}+\sum_{n=0}^\infty \mathrm{e}^{\pi i n s}\big(2A^{\text{toy}}_{T^{-2n}} -\mathrm{e}^{-\pi i u}A^{\text{toy}}_{\tilde{T}^{-2}T^{-2n}}-\mathrm{e}^{-\pi i t}A^{\text{toy}}_{T^{-2n-2}\tilde{T}^{2}} \big) \ . 
\end{align}
Notice that the identity term appears twice in the infinite sum and is subtracted again once in order not to overcount.
We used also that the first and last saddles in \eqref{eq:elementary saddle list} are related by conjugation with the matrix $U$ and thus give the same contribution as discussed in Section~\ref{subsec:elementary saddles}. The relative sign choice between the terms in \eqref{eq: im toy saddle 3 terms} is fixed by the orientation of the steepest descent contour.\footnote{A simple way to confirm this is to put $\theta=\frac{\pi}{2}$, where $-\tau_\theta^*=\tau_\theta+1$ and the different terms recombine into the first term that runs over half-integer $n$. Because of the phases coming from $\vartheta_3(\tau)$, this requires a relative minus sign.}
For each of the three terms, the only $n$-dependence in the saddle point value is via $(\Im \tau)^{-2}$ in \eqref{eq:Agamma toy expression} and one can perform the sum over $n$ somewhat explicitly via the Lerch transcendent function $\Phi(\mathrm{e}^{\pi i s},2,\alpha)$ defined as
\be 
\Phi(z,2,\alpha)=\sum_{n=0}^\infty \frac{z^n}{(n+\alpha)^2}\ .
\ee
One finds the final expression
\begin{multline} 
\mathcal{A}^\text{toy}\sim \frac{2i \, \mathrm{e}^{-s \mathcal{S}(\theta)}\tan(\frac{\theta}{2})^2}{s \pi}\bigg(\frac{\mathrm{e}^{-\pi i u}\Phi(\mathrm{e}^{\pi i s},2,-\tau_\theta)}{\vartheta_3(\tau_\theta)^8}+\frac{\mathrm{e}^{-\pi i t}\Phi(\mathrm{e}^{\pi i s},2,1+\tau^*_\theta)}{\vartheta_3(-\tau^*_\theta)^8}\\
-\frac{2\,  \Phi(\mathrm{e}^{\pi i s},2,-i \Im(\tau_\theta))+\Im (\tau_\theta)^{-2}}{|\vartheta_3(\tau_\theta)|^8}\bigg)\ .
\end{multline}
The imaginary part can be evaluated even more explicitly via Equation~\eqref{eq: toy integral im part formula} and the identity
\be 
\sum_{n \in \ZZ} \frac{\mathrm{e}^{\pi i n s}}{(\alpha-2n)^2}=-\frac{\pi^2\, \mathrm{e}^{\pi i \alpha\left(\{\frac{s}{2}\} -1\right)}\left((\mathrm{e}^{\pi i \alpha}-1)\{\frac{s}{2}\}-\mathrm{e}^{\pi i \alpha}\right)}{4 \sin^2\left(\frac{\pi \alpha}{2}\right)}\ ,
\ee
which can be demonstrated via the Sommerfeld-Watson transform. Here $\{x\}$ denotes the fractional part of $x$. For example, one can derive the compact result
\begin{align}
    \Im \mathcal{A}^{\text{toy}}\big(s,\tfrac{\pi}{2}\big)\sim (2\pi)^7\, 2^{-s}s^{-1}\mathrm{e}^{-2\pi \{\frac{s}{4}\}}  \frac{(\mathrm{e}^{2\pi}-1)\{\frac{s}{4}\}+1}{\Gamma\left(\frac{1}{4}\right)^8\sinh(\pi)^2} \ ,
\end{align}
where the special values of the theta-functions listed in \eqref{eq:special values modular functions} where used.
Clearly, the $s$ dependence is still fairly non-trivial in the large $s$-limit. We display the comparison between saddle point approximation and exact evaluation for both the real and imaginary part in Figure~\ref{fig: plots toy} and find agreement.

\section{\label{sec:saddle bootstrap}Saddle multiplicity bootstrap}
At this point, we know that the string amplitude can be written in the high energy limit as
\be 
A(s,\theta)\sim \sum_{\gamma \in \Gamma(2)/(\gamma \sim U \gamma U)} M_\gamma(s,\theta)\, A_\gamma(s,\theta) \ .\label{eq:amplitude sum over saddles}
\ee
Since $\gamma$ and $U\gamma U$ give identical saddle point contributions, we quotient by this identification.
The multiplicities are themselves sums of multiplicities over the different braidings,
\be 
M_\gamma(s,\theta)=\sum_{\begin{subarray}{c} n \in 2\ZZ+\frac{1}{2}\gamma_{1,2} \\ m \in 2\ZZ+\frac{1}{2}\gamma_{2,1} \end{subarray}} M_{\gamma,n,m} \mathrm{e}^{\pi i n s+\pi i m u}\ . \label{eq:M infinite sum}
\ee
As mentioned before, the multiplicities are integer in our normalization, $M_{\gamma,n,m} \in \ZZ$. Our goal is to determine them completely.

\subsection{Analytic continuation in \texorpdfstring{$s$}{s}} \label{subsec:analytic continuation}
In order to determine the saddle point multiplicities, it is useful to allow $s$ to become complex, while keeping the angle $\theta$ real. Defining this analytic continuation is somewhat non-trivial. For this purpose, let us go back to the integral representation of the amplitude \eqref{eq:string integral}. Recall that $t=-s \sin^2(\frac{\theta}{2})$. We continue to assume that $\theta \in (0,\pi)$, but allow $s$ to become complex. 

\paragraph{Analytic continuation to $\Im s \ge 0$.} Let us begin by considering the simpler case of analytically continuing to $\Im s \ge 0$. Consider the behaviour of the integrand of the string amplitude \eqref{eq:string integral} as $\Im \tau \to \infty$. Let us write $z_i=x_i+i \Im \tau y_i$ with $x_i$ and $y_i$ real. Then $\Im z_{ij}=y_{ij}\, \Im \tau$. The dominating term arises from the exponential in \eqref{eq:string integral}, as well as the leading term in the $q$-expasion of the theta-functions, which leads to
\be 
\mathrm{e}^{-2\pi s \Im \tau\, \text{Trop}(y_i,\theta)}=\exp\bigg(2\pi \Im \tau \sum_{1 \le j<i \le 4} s_{ij} \{y_{ij}\}(1-\{y_{ij}\}) \bigg)\ ,
\ee
with $\{y\}$ denoting the fractional part of $y$.
Subleading terms are suppressed by powers of $\mathrm{e}^{2\pi i \tau}$ or $\mathrm{e}^{-2\pi i \bar{\tau}}$, arising from the $q$-expansion of the theta-functions. On the subregion $\text{Trop}_{<0}$ in the moduli space of the punctures defined by
$\mathrm{Trop}(y_i,\theta)<0$, the integral blows up as we approach the cusp and gets regulated by the $i \varepsilon$ prescription. These regions are discussed in \cite[Section 4.6]{Eberhardt:2022zay}. On the complementary region $\text{Trop}_{>0}$, we can omit the $i \varepsilon$ prescription and integrate over the original fundamental domain. Thus, we can write also the convergent representation of the amplitude
\begin{multline} 
A=\bigg(\int_{\mathcal{F}_{i \varepsilon}} \frac{\mathrm{d}^2 \tau}{(\Im \tau)^2} \int_{\text{Trop}_{<0}} \prod_{i=1}^3 \d x_i\, \d y_i+\int_{\mathcal{F}} \frac{\mathrm{d}^2 \tau}{(\Im \tau)^2} \int_{\text{Trop}_{>0}} \prod_{i=1}^3 \d x_i\, \d y_i\bigg)\\
\times \prod_{1 \le j<i \le 4} \big|\vartheta_1(x_{ij}+\Im \tau \, y_{ij} \, | \, \tau) \big|^{-2s_{ij}} \mathrm{e}^{2\pi i s_{ij} (\Im y_{ij})^2} \ .
\end{multline}
This expression is suitable for analytic continuation to $\Im s>0$. Indeed, the integral over the truncated fundamental domain $\mathcal{F}_L =\{ \tau \in \mathcal{F}\, | \, \Im \tau \le L\}$ can be straightforwardly analytically continued since the integration region is compact. On the region of the contour where $\Trop>0$, the integral over the full fundamental domain is convergent, even when taking $s$ into the upper half plane. For the part of the $i \varepsilon$ contour that goes into the complex plane, we set $\Im \tau=L+i \zeta$ with $\zeta>0$. The integrand is then of the form
\be 
\mathrm{e}^{-2\pi s (L+i \zeta) \, \text{Trop}(y_i,\theta)}+\text{subleading}\ .
\ee
All corrections arising from the $q$-expansion lead to oscillatory terms in $\zeta$. The main point is now that $\zeta \, \text{Trop}(y_i,\theta)<0$, which means that the integral converges as long as $\Im s\ge 0$. In particular, when $s$ is purely imaginary with $\Im s \ge 0$, the $i \varepsilon$-prescription is superfluous and we can simply integrate over the fundamental domain, which converges. Thus for $s \in i \RR_{>0}$, we have
\be
A = \int_{\mathcal{F}} \frac{\d^2 \tau}{(\Im \tau)^5} \int_{\mathbb{T}^2} \prod_{j=1}^3 \d^2 z_j\ \prod_{1 \le j<i \le 4} |\vartheta_1(z_{ij} | \tau)|^{-2s_{ij}}\, \mathrm{e}^{\frac{2\pi s_{ij} (\Im z_{ij})^2}{\Im \tau}}\ . \label{eq:analytic continuation Im s>0}
\ee
A similar analytic continuation of the integral was also discussed in \cite{DHoker:1994gnm}.

\paragraph{Analytic continuation to $\Im s \le 0$.} We are also interested in the analytic continuation of the amplitude to $\Im s \le 0$. For the opposite choice of the $i \varepsilon$ prescription, this works exactly the same as above, but it requires us to move to the second sheet for our choice of $i \varepsilon$ prescription. For this purpose, it suffices to discuss the analytic continuation of the discontinuity of the amplitude. For $\Im s=0$, the discontinuity agrees with the imaginary part.

Such an analytic continuation is by far not unique: there are branch cuts on the real axis at every massive threshold of the form $s=(\sqrt{m_\D}+\sqrt{m_\mathrm{U}})^2$, with $m_\D,\,m_\mathrm{U} \in \ZZ_{\ge 0}$ parametrizing the mass level of the internal states (i.e. $m_\mathrm{D}$ and $m_{\mathrm{U}}$ measure the mass squared of each of the exchanged states). When crossing to a second sheet, one passes through a finite number of branch cuts associated to $m_\mathrm{D}$ and $m_\mathrm{U}$. On the level of the integral representation, this analytic continuation goes as follows. One writes schematically for $s >0$
\be 
\Im A=\frac{1}{2i}\int_{\longrightarrow} \frac{\d^2 \tau}{(\Im \tau)^2} \int \prod_{i=1}^3 \d x_i\, \d y_i \sum_{\sqrt{m_\mathrm{D}}+\sqrt{m_\mathrm{U}} \le \sqrt{s}} \mathcal{I}_{m_\mathrm{D},m_\mathrm{U}}\ .
\ee
Here, $\mathcal{I}_{m_\mathrm{D},m_\mathrm{U}}$ is the part of the integrand obtained by $q$-expansion that contributes to the corresponding discontinuity. $\longrightarrow$ denotes the contour over $\tau$, where we integrate $\Im \tau$ in an imaginary direction. The contour is the difference $\mathcal{F}_{i \varepsilon}-\mathcal{F}_{-i \varepsilon}$. We then restrict the integral over $x_i$ and $y_i$ to the region where $\text{Trop}<0$, since the remaining part of the integral decays as $\Im \tau \to \infty$ and will thus not contribute to the integral. One can then deform the contour over $\Im \tau$. Since all essential singularities of the theta-functions on the real axis are removed because we truncated the $q$-expansion, we can deform the contour over $\Im \tau$ into the lower half-plane, as long as we avoid the explicit pole at $\Im \tau =0$. We thus obtain the equivalent formula for $\Im A$, which is suitable for analytic continuation,
\be 
\Im A=\frac{1}{2i}\int_{\tikz[baseline=-0.5ex] 
{\draw[<-] (.5,.1) to (0,.1);
\draw (.5,-.1) to (0,-.1);
\draw (0,.1) arc (90:270:.1);
}} \frac{\d^2 \tau}{(\Im \tau)^2} \int_{\text{Trop}<0} \prod_{i=1}^3 \d x_i\, \d y_i \sum_{\sqrt{m_\mathrm{D}}+\sqrt{m_\mathrm{U}} \le \sqrt{s}} \mathcal{I}_{m_\D,m_\mathrm{U}}\ ,
\ee
where the contour of $\Im \tau$ is a Hankel contour, i.e.\ runs from $-iL+\infty$ to $-i L$, then around the origin to $i L$ and then back to $i L+\infty$ as indicated by the arrow. The sum over $m_\D$ and $m_\mathrm{U}$ is finite, depending on the sheet which one wants to analytically continue. 

We are now interested in the behaviour of the amplitude if we first take the large $s$ limit and then analytically continue to $\Im s \le 0$, since this will have an imprint in the saddle point expression. We have
\be 
\frac{1}{s}\log \sum_{\sqrt{m_\mathrm{D}}+\sqrt{m_\mathrm{U}} \le \sqrt{s}} \mathcal{I}_{m_\D,m_\mathrm{U}} \overset{s \to \infty}{\longrightarrow} F(\boldsymbol{m})
\ee
for some function $F(\boldsymbol{m})$ of the moduli. For large enough $\Im \tau$, the sum is rapidly convergent and the upper bound on $m_\mathrm{D}$ and $m_\mathrm{U}$ is irrelevant. Thus the limiting function is just the original integrand. For small (or even negative) $\Im \tau$, the sum is dominated by the last few terms in the sum and tends to a \emph{different} limiting function.
Thus $F(\boldsymbol{m})$ is only piecewise analytic. As a simple example of such a phenomenon, consider the function $f(y)=(1-y)^{-s}=\sum_{k \ge 0} \binom{s+k-1}{k} y^k$. Let us truncate the sum at $k=\lfloor s \rfloor $. Then for $y<\frac{1}{2}$, the summands decay and we recover the original function in the limit $s \to \infty$. For $y>\frac{1}{2}$, the summands grow too quickly and the sum is dominated by the last summand. One obtains
\be 
\frac{1}{s} \log \sum_{k=0}^{\lfloor s \rfloor } {s+k-1 \choose k} y^k \overset{s \to \infty}{\longrightarrow} \begin{cases}
    -\log(1-y)\ , \qquad y<\frac{1}{2}\ , \\
    \log(4y)\ , \qquad y>\frac{1}{2}\ .
\end{cases}
\ee
This means that up to subexponential corrections, the imaginary part in the large-$s$ limit, can be written again as
\be 
\Im A\sim \frac{1}{2i}\int_{\tikz[baseline=-0.5ex] 
{\draw[<-] (.5,.1) to (0,.1);
\draw (.5,-.1) to (0,-.1);
\draw (0,.1) arc (90:270:.1);
}} \frac{\d^2 \tau}{(\Im \tau)^2} \int_{\text{Trop}<0} \prod_{i=1}^3 \d x_i\, \d y_i \, F(\boldsymbol{m})^{-s}\ ,
\ee
which is amenable to saddle point evaluation. In particular, it is now straightforward to analytically continue it to purely imaginary $s$, for which we can write
\be 
\Im A\sim \frac{1}{2i}\int_{\, \tikz[baseline=.5ex, rotate=90] 
{\draw[<-] (.4,.1) to (0,.1);
\draw (.4,-.1) to (0,-.1);
\draw (0,.1) arc (90:270:.1);
}} \frac{\d^2 \tau}{(\Im \tau)^2} \int_{\text{Trop}<0} \prod_{i=1}^3 \d x_i\, \d y_i \, F(\boldsymbol{m})^{-s}\ , \label{eq:analytic continuation Ims<0}
\ee
where the contour is now rotated by 90 degrees.
Since $F$ is a sum of real terms, it is real and we obtain an oscillatory integral, which is again bounded by a constant as $\Im s$ becomes large. 

\paragraph{Boundedness in the complex plane.} We now use the previous analytic continuation to bound the behaviour of the amplitude in the complex plane. 
In the extreme case where $s$ is purely imaginary, for both $\Im s>0$ and $\Im s<0$, the analytically continued integral is purely oscillatory and thus bounded by a constant in $s$.\footnote{In the case $\Im s<0$, we strictly speaking only derived that the amplitude grows subexponentially, but it will follow from the saddle point structure that it is in fact totally bounded.} Thus, it follows that for $s \in i \RR$ and $\theta \in (0,\pi)$ that
\be 
|A(s,\theta)| \le C(\theta)\label{eq:boundedness property imaginary line}
\ee
for some constant $C(\theta)$.
The same property is true for $s$ in the first quadrant, which is most easily established using the Phragm\'en-Lindel\"of principle. The saddle-point expression \eqref{eq:amplitude sum over saddles} satisfies for $\varepsilon>0$
\be 
s \in \RR_{\ge 0}+i \varepsilon:\quad \big|s^4\, \mathrm{e}^{s\, \Re S_*(\theta)}\, A(s,\theta)\big| <C(\varepsilon, \theta)\ .
\ee
The constant $C(\varepsilon,\theta)$ diverges as $\varepsilon \to 0$ because of the double poles at the integers. Thus the combination $\tilde{A}(s,\theta)=\mathrm{e}^{s \Re S_*(\theta)} A(s,\theta)$
is bounded on the rays $s \in \RR_{\ge 0}+i \varepsilon$ and $s \in i \RR_{\ge 0}$. Furthermore, in other directions of the complex plane, $A(s,\theta)$ can grow at most exponentially as a function of $|s|$, which follows from saddle point evaluation of the integral. Thus the Phragm\'en-Lindel\"of principle applies. The same argument works in the lower half plane. Thus we conclude that for $\Re s \ge 0$ and $|\Im s| \ge \varepsilon$
\be 
\big| \mathrm{e}^{s \, \Re S_*(\theta)}\, A(s,\theta)\big| < C(\varepsilon,\theta)\ . \label{eq:boundedness property}
\ee
We will use this bound to put further constrain multiplicities of saddles. We also think \eqref{eq:boundedness property} is of more general interest in the S-matrix bootstrap program.

In the simpler case of $\Im s \to +\infty$, the saddle point expression should tends to the integral \eqref{eq:analytic continuation Im s>0}, which in turn is dominated by the real Gross--Mende saddle point with multiplicity 1. 

\subsection{Further constraints and assumptions}
\paragraph{Stokes phenomena.} In principle, the saddle point multiplicities can jump discontinuously as a function of $\theta$. To decide whether such Stokes phenomena happen we would need to construct explicitly the steepest descent contour which is beyond our capabilities. In all numerical data that we computed, no Stokes phenomenon is visible as a function of $\theta$. Furthermore, one physically expects that the amplitude has a smooth asymptotics as a function of the angle $\theta$. Therefore we assume that such Stokes phenomena are absent.\footnote{Let us emphasize that we restrict to the $s$-channel. Of course, the saddle-point multiplicities in the $t$ and $u$ channel or unphysical channels are expected to be different.}

\paragraph{Trigonometric functions.}
Let us begin by considering $M_\gamma(s,\theta)$ given in \eqref{eq:M infinite sum} for fixed $\gamma$. The sum over $n$ and $m$ is very similar to the tree-level sum over saddles \eqref{eq:tree level trigonometric geometric series}. In the tree-level case, the amplitude has poles at integer $s \in \ZZ$ because of string resonances. Likewise, the one-loop amplitude has a double pole at integer $s \in \ZZ$ because there are two propagators that can go on-shell (in field theory, it would correspond to the bubble insertion on the propagator). This double pole has to be produced by the sum over multiplicities. Notice that the series
\be 
\frac{1}{4\sin^2(\pi s)}=-\sum_{n=1}^\infty n\, \mathrm{e}^{-2\pi i n s}
\ee
produces the correct pole. Thus we reorganize the sum over saddles for fixed $\gamma$ into
\be 
M_{\gamma}(s,\theta)=\frac{1}{4 \sin^2(\pi s)}\sum_{\begin{subarray}{c} n \in 2\ZZ+\frac{1}{2}\gamma_{1,2} \\ m \in 2\ZZ+\frac{1}{2}\gamma_{2,1} \end{subarray}}\tilde{M}_{\gamma,n,m} \mathrm{e}^{\pi i n s+\pi i m u}\ . \label{eq:multiplicities rewritten}
\ee
Since we do not expect other poles on physical grounds, the sum over $n$ and $m$ is expected to be \emph{finite} in this second representation.

\paragraph{Implication for the saddle point multiplicities.} Let us return to the saddle point expression \eqref{eq:amplitude sum over saddles} and investigate what the boundedness property \eqref{eq:boundedness property} implies for the multiplicities. For this, we have to assume that the saddle-point expression is valid also for complex $s$, i.e.\ that there are no Stokes phenomena as $s$ becomes complex. As long as $\arg s <\frac{\pi}{2}$, subleading saddles such as those discussed in Section~\ref{subsec:saddle hunting} are still subleading and the asymptotic equality is expected to hold.

Each individual $A_{\gamma}(s,\theta)$ satisfies the boundedness property. However, the multiplicities in the form \eqref{eq:multiplicities rewritten} can violate it if $|n+m \frac{u}{s}|=|n-m \cos^2(\frac{\theta}{2})|>2$. Thus, this seems to suggest that $\tilde{M}_{\gamma,n,m}=0$ if $|n-m \cos^2(\frac{\theta}{2})|>2$ for some choice of $\theta \in (0,\pi)$. This turns out to be not quite true. The caveat is that each individual in \eqref{eq:amplitude sum over saddles} can violate this condition, but after summing over $\gamma \in \Gamma(2)$, boundedness can be restored. In fact, we saw that this occurred in the toy integral, and we expect it to also happen in this case. 

\subsection{Tail sum over saddles} \label{subsec:tail sum}
The infinite sum over saddles can lead to non-trivial analytic features, such as the poles $s \in \ZZ$ in \eqref{eq:multiplicities rewritten}. Similarly, there are new analytic features arising from the tail of the infinite sum over $\gamma$ in \eqref{eq:amplitude sum over saddles}. We already know a lot about this tail. As argued in \eqref{subsec:elementary saddles}, the steepest descent contour in $\tau$ reduces to the steepest descent contour of the toy integral in this tail. In particular, this tail is controlled by the elementary saddles and the relative phases between the saddles are the same as in the toy integral. Thus, we obtain
\begin{multline} 
A_\text{tail}(s,\theta)=M_\text{tail}(s,\theta)\sum_{n=0}^\infty \mathrm{e}^{\pi i n s}\big(2A_{T^{-2n}}(s,\theta)  \\ +\mathrm{e}^{-\pi i u}A_{\tilde{T}^{-2}T^{-2n}}(s,\theta)+\mathrm{e}^{-\pi i t}A_{T^{-2n-2}\tilde{T}^{2}}(s,\theta) \big)\ . \label{eq:tail sum over saddles}
\end{multline}
Compare also with the expression \eqref{eq: im toy saddle 3 terms} of the toy integral.\footnote{In that case, there was a relative minus sign between the terms. One can check by the same argument that in this case, we should have a relative plus sign between these terms.}
Here, $M_\text{tail}(s,\theta)$ is the joint multiplicity for all these saddles, which is of the same form as in \eqref{eq:multiplicities rewritten} with $\gamma_{1,2}\equiv \gamma_{2,1} \equiv 0 \bmod 4$. We included a factor of 2 in front of $A_{T^{-2n}}(s,\theta)$, since it is equivalent to the saddle $\tilde{T}^{-2}T^{-2n}\tilde{T}^2$. We then have
\be 
A(s,\theta) \sim A_\text{tail}(s,\theta)+\sum_{\gamma \in \Gamma(2)/(\gamma \sim U \gamma U)} M'_\gamma(s,\theta) A_\gamma(s,\theta)\ , \label{eq:tail sum subtraction}
\ee
but contrary to \eqref{eq:amplitude sum over saddles} only finitely many multiplicities $M'_\gamma(s,\theta)$ should be non-zero.

\paragraph{Analytic properties.} Let us analyze the analytic properties of $A_\text{tail}(s,\theta)$. As we already remarked above, each individual term in the tail sum over saddles \eqref{eq:tail sum over saddles} violates the boundedness property (for large enough $n$) and we have to make sure that the sum restores it. Let's begin with the sum \be 
\sum_{n=0}^\infty \mathrm{e}^{\pi i n s}A_{T^{-2n}}(s,\theta)
\ee
with $A_\gamma(s,\theta)$ given in \eqref{eq:Agamma expression}. For the purpose of computing this sum, we notice that all theta functions as well as the second Eisenstein series appearing in \eqref{eq:determinant Hessian} are invariant under $\tau \to \tau+2$. The only piece that is not invariant is $\Im \tau$. Thus, the Hessian is of the form $C_0\, (\Im \tau)^{-6}\prod_{j=1}^3((\Im \tau)^2-C_j)$ with $C_0$ and $C_j$ independent of $n$. The sum is thus of the form
\begin{align} 
\sum_{n=0}^\infty \mathrm{e}^{\pi i n s}A_{T^{-2n}}(s,\theta)&=\frac{C_0' \mathrm{e}^{-s \Re_*(s,\theta)}}{s^4} \sum_{n=0}^\infty \frac{\mathrm{e}^{\pi i n s}}{(\Im \tau)^2 \prod_{j=1}^3 \sqrt{C_j-(\Im \tau)^2}}\bigg|_{\tau=\tau_\theta-2n,\bar{\tau}=\tau_\theta^*} \\
&=\frac{C_0' \mathrm{e}^{-s \Re_*(s,\theta)}}{s^4} \sum_{n=0}^\infty \frac{\mathrm{e}^{\pi i n s}}{(\Im \tau_\theta+i n)^2 \prod_{j=1}^3 \sqrt{C_j-(\Im \tau_\theta+i n)^2}} \ . \label{eq:tail sum over saddles 1}
\end{align}
Clearly, this function remains bounded for $s$ in the upper half plane, since the sum will be dominated by the $n=0$ term. To investigate the behaviour for $s$ in the lower half plane, we first have to define the analytic continuation. This can be done with the help of the Sommerfeld-Watson trick. Let us assume that $0 \le \Re s \le 2$, since the infinite sum is clearly invariant under $s \to s+2$. We can write the sum as a contour integral enclosing all non-negative integers: 
\begin{align}
    &\sum_{n=0}^\infty \frac{\mathrm{e}^{\pi i n s}}{(\Im \tau_\theta+i n)^2 \prod_{j=1}^3 \sqrt{C_j-(\Im \tau_\theta+i n)^2}}\nonumber\\
    &\qquad=\oint_{\ZZ_{\ge 0}} \d z\, \frac{\mathrm{e}^{\pi i z s}}{(\mathrm{e}^{2\pi i z}-1)(\Im \tau_\theta+i z)^2\prod_{j=1}^3 \sqrt{C_j-(\Im \tau_\theta+i z)^2}} \\
    &\qquad=-\oint_{\ZZ_{< 0}\cup \{\text{poles, cuts}\}} \d z\, \frac{\mathrm{e}^{\pi i z s}}{(\mathrm{e}^{2\pi i z}-1)(\Im \tau_\theta+i z)^2\prod_{j=1}^3 \sqrt{C_j-(\Im \tau_\theta+i z)^2}}\ ,
\end{align}
where from the second line to the third, we use that the integrand tends to zero as $|z| \to \infty$ because $0 \le \Re s \le 2$. This allows us to rewrite the contour integral in terms of all other singularities in the complex plane, which are located at negative integers, the explicit pole at $z=-i \Im \tau_\theta$ and the cuts starting from the branch points at $z=i\, \Im \tau_\theta\pm i\sqrt{C_j}$. The main point is that this new representation converges now for $\Im s \le 0$ and thus defines the analytic continuation into the lower half plane. By the same argument as before, it is bounded for $s$ in the lower half plane. The contributions from the other singularities in the $z$ plane are polynomially bounded. In fact a singularity of the form $(z-z_*)^{-p}$ leads to a contribution that grows like $s^{p-1}$. At worst, we can get a fifth order pole if all singularities collide, which means that the function is bounded by $s^4$, exactly compensating the $s^{-4}$ prefactor present in \eqref{eq:tail sum over saddles 1}. Thus, for $\Re s\ge 0$
\be 
\Big|\mathrm{e}^{s \Re S_*(\theta)}\sum_{n=0}^\infty \mathrm{e}^{\pi i n s}A_{T^{-2n}}(s,\theta)\Big| \le C\ , 
\ee
which is exactly the boundedness condition \eqref{eq:boundedness property}. We can similarly derive such a bound for the other terms in \eqref{eq:tail sum over saddles}. The boundedness property \eqref{eq:boundedness property} then implies that $|M_\text{tail}(s,\theta)| \le C$ for $\Re s \ge 0$ and $|\Im s| \ge \varepsilon$. This means that the possibilities for $M_\text{tail}(s,\theta)$ are very restricted. They can only take the form.
\be 
M_\text{tail}(s,\theta)=\frac{m_1\, \mathrm{e}^{2\pi i s}+m_2 \, \mathrm{e}^{-2\pi i s}+m_3\, \mathrm{e}^{2\pi i t}+m_4 \mathrm{e}^{-2\pi i t}+m_3\, \mathrm{e}^{2\pi i u}+m_4 \mathrm{e}^{-2\pi i u}+m_5}{4 \sin^2(\pi s)}\ , \label{eq:tail multiplicity}
\ee
where we already imposed that the result has to be crossing symmetric. All other combinations would violate the boundedness property for some angle $\theta \in (0,\pi)$. The remaining multiplicities in \eqref{eq:tail sum subtraction} also have to be of this form since there are only finitely many of them and they cannot resum, except that they are not necessarily crossing symmetric, but are related to the multiplicities of the saddle labelled by $C\gamma C$ as discussed in Section~\ref{subsec:systematics}.

\subsection{Thresholds and non-analyticities}
\paragraph{Comparison with the massive thresholds.} As mentioned before, there is an obvious source of branch points in the amplitude from the massive thresholds, which, are located at $s=s_0=(\sqrt{m_\D}+\sqrt{m_\mathrm{U}})^2$.

\begin{figure}
    \centering
    \includegraphics[width=0.7\textwidth]{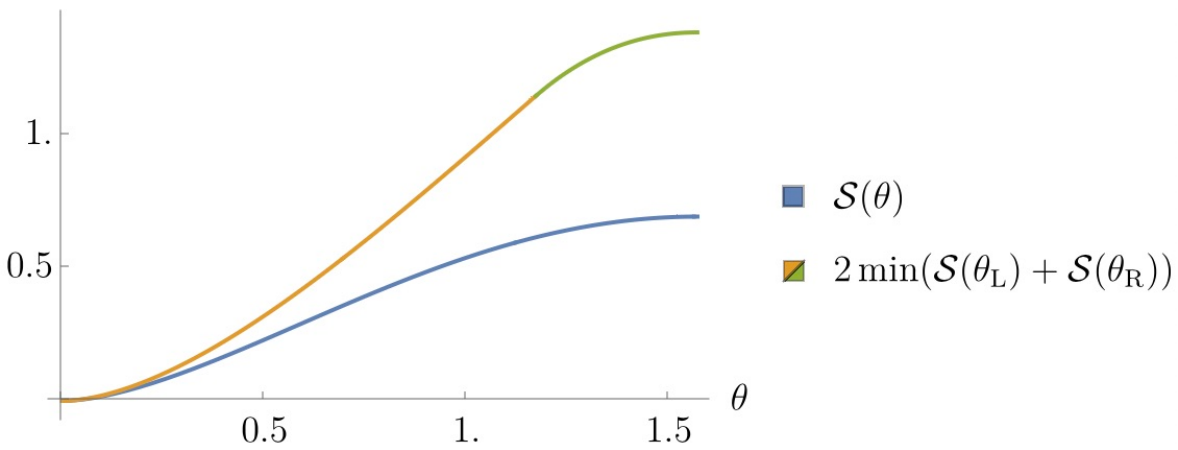}
    \caption{The fall-off of a contribution from a threshold compared to the fall-off of the whole amplitude. At the transition from the yellow to the green part of the curve, the location of the minimum changes qualitatively as described in the text.}
    \label{fig:Baikov fall off}
\end{figure}

Every individual branch cut leads to a discontinuity whose detailed local behaviour depends on the value of $m_\D$ and $m_\mathrm{U}$: For generic $m_\D$ and $m_\mathrm{U}$, the local behaviour is $\sim (s-s_0)^{\frac{7}{2}}$. When either $m_\D=0$ or $m_\mathrm{U}=0$, we get the local behaviour $\sim (s-s_0)^4 \log(s-s_0)$, while for $m_\D=m_\mathrm{U}=0$, we get the local behaviour of the supergravity amplitude (which involves a dilog). The coefficient of these discontinuities decay in the $s \to \infty$ limit as follows. Up to polynomial corrections coming from the sum over polarizations (the $Q$-polynomial in the Baikov formula, see \cite[Equation~(4.6)]{Baccianti:2025whd}), the exponential fall-off of the contribution from the threshold is the product of the fall-offs of the two tree-level amplitudes taking part in the unitarity cut, i.e.
\be
\sim \mathrm{e}^{-2s (\mathcal{S}(\theta_\L)+\mathcal{S}(\theta_\R))}\,,
\ee
where $\mathcal{S}(\theta)$ is the on-shell action from \eqref{eq:real part one loop action} and $\theta_\text{L}$ and $\theta_\text{R}$ are the scattering angles of the two subprocesses. For on-shell kinematics, they are bounded in terms of the total scattering angle as
\be
|\theta_\L-\theta_\R| \le \theta \le \min(\theta_\L+\theta_\R,2\pi-\theta_\L-\theta_\R)\, .
\ee
Since the unitarity formula integrates over all possible values of $\theta_\L$ and $\theta_\R$, the exponential fall-off is determined by the choice of $\theta_\L$ and $\theta_\R$ for which $2\mathcal{S}(\theta_\L)+2\mathcal{S}(\theta_\R)$ is minimal. For $0<\theta<1.17056$\footnote{The upper bound is the unique solution of the equation $\cos\frac{\theta}{2}\,  \log \cot \frac{\theta}{4}=1$.}, this solution is attained for $\theta_\L=\theta_\R=\frac{\theta}{2}$. For $1.17056<\theta<\frac{\pi}{2}$, it is attained for $\theta_\L+\theta_\R=\theta$, but there exists a smaller minimum of $\mathcal{S}(\theta_\L)+\mathcal{S}(\theta-\theta_\L)$ in the interval $\theta_\L \in [0,\theta]$. The resulting minimum of $2\mathcal{S}(\theta_\text{L})+2\mathcal{S}(\theta_\text{R})$ is plotted in Fig.~\ref{fig:Baikov fall off}. 
Importantly, 
\be 
\min_{\theta_\L,\theta_\R}\, \big(2\mathcal{S}(\theta_\text{L})+2\mathcal{S}(\theta_\text{R})\big)>\mathcal{S}(\theta)\ .
\ee
Thus every individual threshold contribution falls off faster than the whole amplitude.

\begin{figure}
    \centering
    \includegraphics[width=0.5\linewidth]{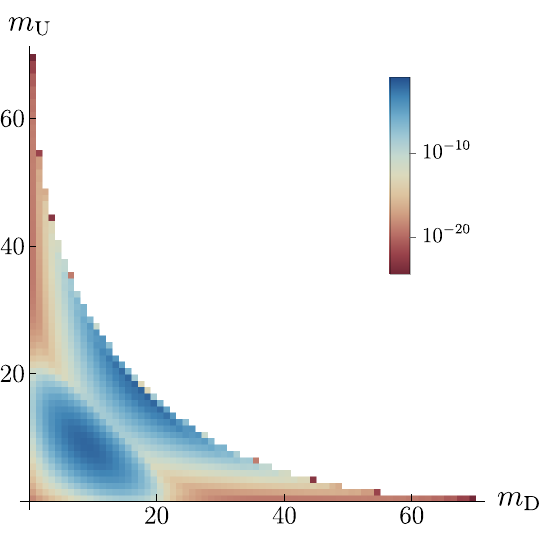}
    \caption{The relative contribution of the different massive thresholds to the full decay width (i.e.\ the coefficient of the double pole) at $s=70$ and $\theta=\frac{\pi}{2}$. In other words, this plot shows the decay width divided into different decay channels. The distribution is not sharply peaked at any particular internal mass level and many different channels have to conspire in order to lead to the expected asymptotic behaviour. }
    \label{fig:mDmU plot}
\end{figure}

It is also interesting to note that this implies that from the point of view of unitarity cuts, the individual threshold contributions have to conspire precisely in order to give the expected exponential fall off of the amplitude. This also means that the asymptotic behaviour originates from many different different internal mass levels and is difficult to analyze directly from the imaginary part of the amplitude. It requires that all the different contributions from different mass levels conspire to result in the predicted asymptotic behaviour. This is illustrated in Figure~\ref{fig:mDmU plot} where we plotted the contributions from the different mass levels to the decay width. For concreteness, we chose $s=70$ and $\theta=\frac{\pi}{2}$ for the plot. The data for this plot is obtained as described in Section~\ref{subsec:numerical data}. See also \cite{Chialva:2003hg} for a similar computation.

This result implies in particular that the thresholds are washed out in the high-energy limit. We shall now demonstrate from the tail sum that this is indeed the case (contrary to the toy integral discussed in Section~\ref{sec:toy integral}), but we instead get new emergent non-analyticities in the high-energy limit at $s \in 2 \ZZ$.

\paragraph{Non-analyticities at $s \in 2 \ZZ$.} The presence of the tail sum leads to surprising non-analyticities in the asymptotics of the amplitude. Let us explain the mechanism. These non-analyticities are generated by the leading large-$n$ behaviour of the tail sum. Let us compute the leading non-analyticity near $s \in 2\ZZ$. For this, we note that
\be 
A_{T^{-2n}}(s,\theta) \xrightarrow{n \to \infty} \frac{(2\pi)^4 \mathrm{e}^{-s \Re S_*(\theta)}}{s^4 (i n)^5} \times \frac{1}{\sqrt{\det H^\infty}} \bigg|_{\tau=\tau_\theta,\bar{\tau}=\tau_\theta^*}\ ,
\ee
where for $H^\infty$, we omit the term $\frac{1}{(\Im \tau)^2}$ in the determinant of the Hessian \eqref{eq:determinant Hessian}, since it disappears for $n \to \infty$. We have essentially the same identities for the other terms in \eqref{eq:tail sum over saddles}. To get the non-analyticities, we only have to look at the sum (we omit the $n=0$ term since it doesn't contribute)
\be 
\sum_{n=1}^\infty  \frac{\mathrm{e}^{\pi i n s}}{n^5}=\sum_{n=1}^\infty \frac{\mathrm{e}^{\pi i n (s-s_0)}}{n^5} \sim \int_1^\infty \d n\,  \frac{\mathrm{e}^{\pi i n (s-s_0)}}{n^5}
\ee
For $s \sim s_0 \in 2 \ZZ$, the sum is very slowly varying with $n$ and can be approximated by an integral. If we then take $(s-s_0) \to \mathrm{e}^{2\pi i}(s-s_0)$, we have to compensate by rotating $n$ in the opposite direction to keep the integral convergent. This will pick up the residue at $n=0$ and leads to a discontinuity of the form
\be 
\Disc_{s=s_0}\int_1^\infty \d n\,  \frac{\mathrm{e}^{\pi i n (s-s_0)}}{n^5} =-2\pi i \Res_{n=0} \frac{\mathrm{e}^{\pi i n (s-s_0)}}{n^5} \ ,
\ee
and so we get the leading local non-analytic behaviour
\be 
\int_1^\infty \d n\,  \frac{\mathrm{e}^{\pi i n (s-s_0)}}{n^5} \sim -\log (s-s_0) \Res_{n=0}\frac{\mathrm{e}^{\pi i n (s-s_0)}}{n^5} =-\frac{\pi^4}{24} (s-s_0)^4 \log(s-s_0)\ .
\ee
Thus the tail sum produces the following leading non-analyticities near $s=s_0 \in 2 \ZZ$,
\begin{multline} 
A_\text{tail}(s,\theta)\sim\frac{2\pi^6i}{s_0^4}\, (s-s_0)^2 \log (s-s_0)\, \mathrm{e}^{-s_0 \Re S_*(\theta)}\bigg(\frac{2}{\sqrt{\det H^\infty}} \bigg|_{\tau=\tau_\theta,\bar{\tau}=\tau_\theta^*}\\
+\frac{\mathrm{e}^{-\pi i u}}{\sqrt{\det H^\infty}} \bigg|_{\tau=\tau_\theta,\bar{\tau}=-\tau_\theta}+\frac{\mathrm{e}^{\pi i u}}{\sqrt{\det H^\infty}} \bigg|_{\tau=-\tau_\theta^*,\bar{\tau}=\tau_\theta^*}\bigg)\ . \label{eq:discontinuities at s in 2Z}
\end{multline}
Here we inserted the values $m_1+m_2+m_5=12$ and $m_3+m_4=0$ of the saddle multiplicities $m_i$ that we will determine below.
The main point of this computation is that we get logarithmic singularities of the form $(s-s_0)^2 \log(s-s_0)$.

\subsection{Numerical data} \label{subsec:numerical data}
At this point, we have exhausted the general constraints on the amplitude. The only way forward short of constructing the actual steepest descent contour is to compare the saddle point prediction with explicit computations of the amplitude. We have several datasets for the amplitude that we will combine.  Let us explain the origin of this data. All of our data is published on \texttt{Zenodo} \cite{Zenodo}. The summary is given in Table~\ref{tab:numerical data}.

\begin{table}
\centering
{\setlength{\tabcolsep}{5.2pt}
\begin{tabular}{c|c|c|c|c|c} 
 quantity & $s$ range & $\theta$ range & method & code & data \\ 
 \hline
 $\Re A(s,\theta)$, $s \in \RR_+$ & $s \lesssim 11$ & $\theta \in \Theta$ & Rademacher & \texttt{StringQMC} \cite{Baccianti:2025whd} & \texttt{ReA/*} \\ 
 $\Im A(s,\theta)$, $s \in \RR_+$ & $s \leq 52$ & $\theta \in \Theta$ & Baikov & \texttt{ImA.nb} & \texttt{ImA/*} \\ 
  $\Re A(s,\theta)$, $s \in \ZZ_+$ & $s \leq 35$ & $\theta \in \CC$ & Rademacher & \texttt{ReDResA.nb} & \texttt{ReDResA/*} \\ 
 $\Im A(s,\theta)$, $s \in \ZZ_+$ & $s \leq 70$ & $\theta \in \CC$ & Baikov & \texttt{ImDResA.nb} & \texttt{ImDResA/*} \\ 
\end{tabular}
}
\caption{\label{tab:numerical data}Summary of the numerical data published on \texttt{Zenodo} \cite{Zenodo}. For each quantity, we give the range of $(s,\theta)$, the method for generating the data, as well as names of files of the code generating the data in the folder \texttt{code/}, and the saved data in the folder \texttt{data/}.}
\end{table}

\paragraph{Imaginary part.} The imaginary part of the amplitude is much simpler to compute than the real part. It can be obtained from unitarity cuts \cite{Eberhardt:2022zay,Banerjee:2024ibt}. In the Baikov representation, the relevant formula is given in \cite[Equation~(3.1)]{Banerjee:2024ibt}. The Baikov representation involves an integral over the on-shell phase space of the intermediate states which can be only done numerically for generic kinematics. We have computed the imaginary part numerically for the range of angles $\sin^2\frac{\theta}{2} \in \{\frac{1}{2},\frac{1}{3},\frac{1}{4},\frac{1}{5},\frac{2}{5},\frac{2}{7},\frac{3}{7}\}$ (in Table~\ref{tab:numerical data}, we denote the set of angles as $\theta \in \Theta$) and $s \le 53$ in steps of $\delta s=\frac{1}{8}$.  The computation gets increasingly costly for large kinematics because of the exponential number of internal states one has to sum over.

For $s \in \ZZ_+$, the amplitude has a double pole. The integral over the on-shell kinematics can be performed analytically for the coefficient of this double pole that we will denote by DRes. The coefficients compute the decay width of the massive string states, when averaged over the degeneracy of the states. The quantity $\Im \DRes_{s=n} A$ is a polynomial in $t$ of order $2n-2$. We computed these decay widths analytically for $s \le 70$ as a function of the angle $\theta$.

\paragraph{Real part.} The real part is much more challenging to compute, especially in the high energy limit, where there are exponentially large cancellations in the original integral and direct numerical evaluation is unfeasible. We instead make use of the evaluation through the Rademacher method that was recently developed \cite{Baccianti:2025gll, Baccianti:2025whd}. To our knowledge, it is the only method that can compute the amplitude reliably for high energies. The Rademacher formula is still complicated. It involves a four-dimensional integral (up from the two-dimensional integral of the Baikov representation), a quadruple sum over internal masses (up from the double sum over the internal masses encountered in the Baikov representation) and an additional infinite sum over Lorentzian singularities of the moduli space $\mathcal{M}_{1,4}^\CC$. The last sum goes over tuples $(a,c;n_\L,n_\D,n_\R)$ with $n_{\L/\D/\R} \in \{0,\dots,c-1\}$. $a \in \{0,\dots,c-1\}$ with $\gcd(a,c)=1$ and $c\in \ZZ_{\ge 1}$. We truncate $c \le c_\text{max}$ for some $c_\text{max}$ so that the sum becomes finite. Convergence in $c_\text{max}$ is fairly slow, but accelerates somewhat when $s$ is close to an integer. We refer to \cite{Baccianti:2025whd} for further details.

We computed the amplitude at the same set of fixed angles $\theta \in \Theta$ as before using the quasi Monte-Carlo code \texttt{StringQMC} \cite{Baccianti:2025whd}. In most cases, the data is numerically stable until $s \lesssim 11$ and we determined it for $c_\text{max} = 3$ in steps of $\delta s=\frac{1}{8}$ for non-integer $s$. For $\theta = \frac{\pi}{2}$, we obtained additional data for $c_\text{max}=5$ up to $s=9.625$. Convergence seems decent and the truncated amplitude has more or less stabilized.

As with the imaginary part, things simplify significantly for integer $s$. In this case, the integrals can be done in closed form in terms of Bessel functions, and the sum over $n_\L$ and $n_\R$ disappears. The sum over $n_\D$ is a classical Gauss sum and can be performed in closed form. This reduction was discussed in detail in \cite{Baccianti:2025whd}. We implemented the resulting formula for $c_\text{max}=10$ and $s \le 35$, which again is valid for all angles.

\subsection{Fourier analysis and results}

It is non-trivial to extract the saddle point multiplicities from the data. In practice, we proceed as follows. First, we multiply the data by the prefactor $s^4\, \mathrm{e}^{s  \mathcal{S}(\theta)} \sin^2(\pi s)$, so that the data is expected to approach an oscillatory behaviour as $s \to \infty$. Since we know the expected phases of the saddles, the contribution from different saddles can be disentangled by performing Fourier analysis over $s$. For this it is convenient to work at a fixed angle $\theta$, where the ratio $-\frac{t}{s}=\frac{p}{q} \in \QQ $ is rational, so that the data has the period $2q$ in $s$, which is the joint period of $\mathrm{e}^{\pi i s}$ and $\mathrm{e}^{\pi i t}$. Therefore, one can multiply the data by the expected exponential and take a moving average over this bin size, which filters out the different contributions. The saddles have different importance, depending on the size of $(\det H_{i,j})^{-\frac{1}{2}}$, given in \eqref{eq:determinant Hessian}. 

\begin{figure}
    \centering
    \begin{minipage}{.485\textwidth}
    \includegraphics[width=\textwidth]{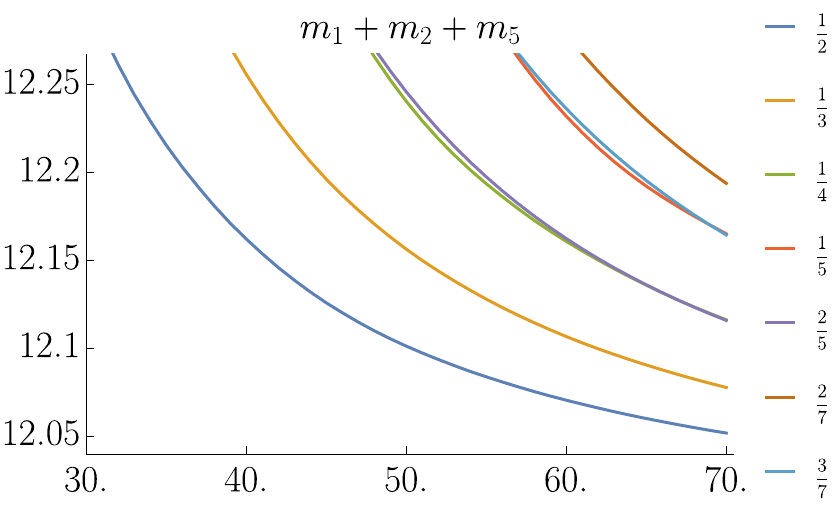}
    \end{minipage}
    \begin{minipage}{.485\textwidth}
    \includegraphics[width=\textwidth]{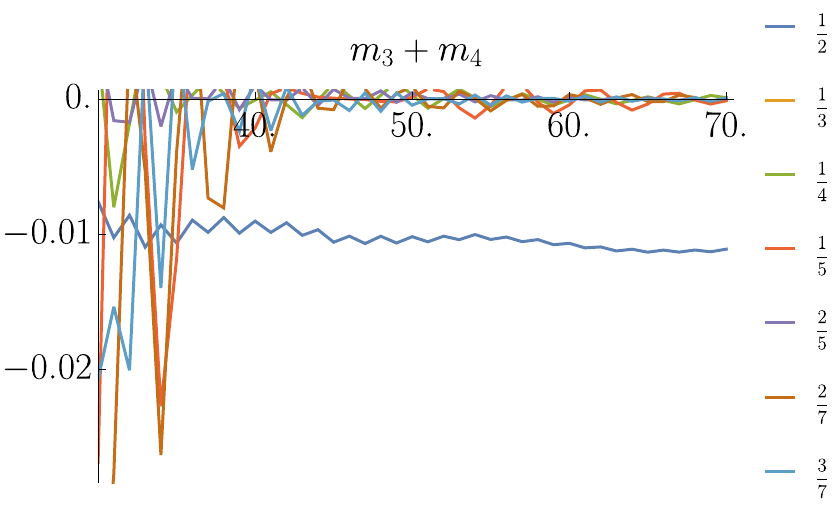}
    \end{minipage}
    \begin{minipage}{.485\textwidth}
    \includegraphics[width=\textwidth]{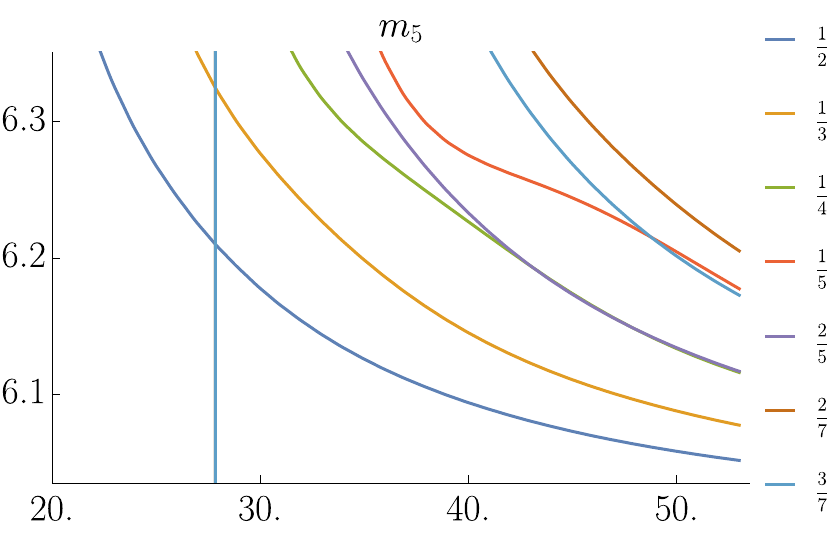}
    \end{minipage}
    \begin{minipage}{.485\textwidth}
    \includegraphics[width=\textwidth]{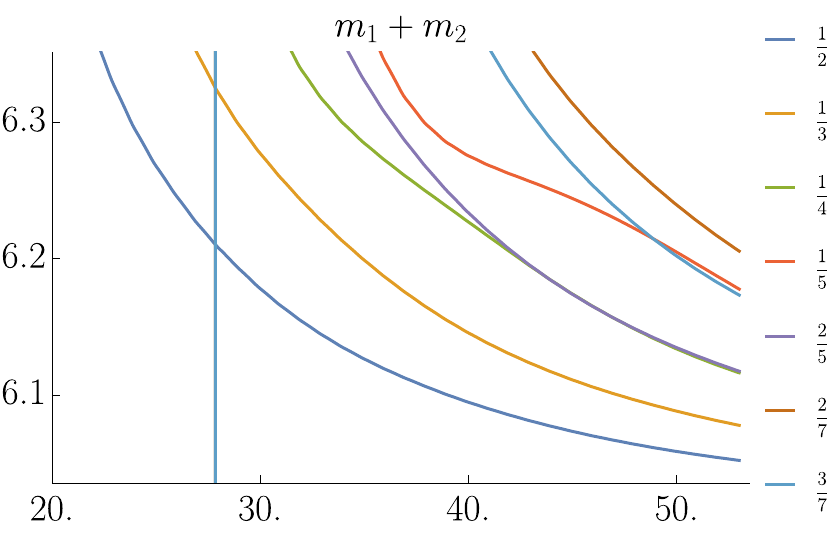}
    \end{minipage}
    \begin{minipage}{.485\textwidth}
    \includegraphics[width=\textwidth]{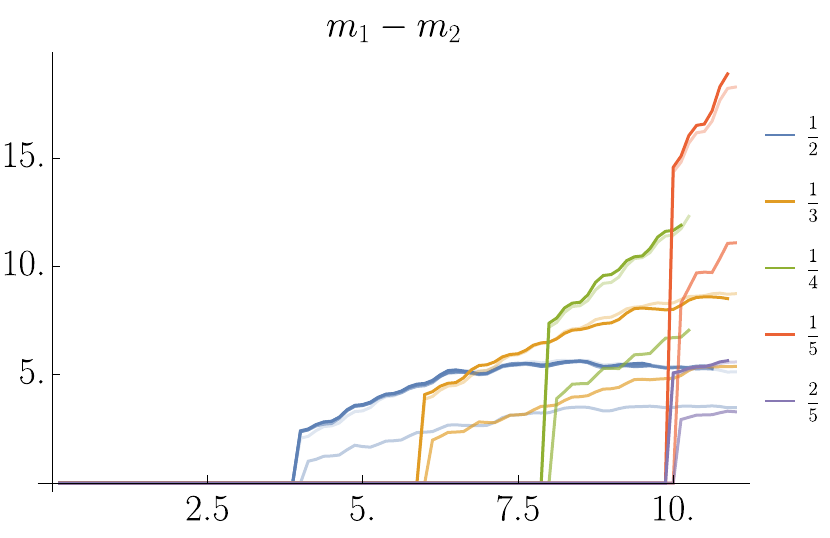}
    \end{minipage}
    \begin{minipage}{.485\textwidth}
    \includegraphics[width=\textwidth]{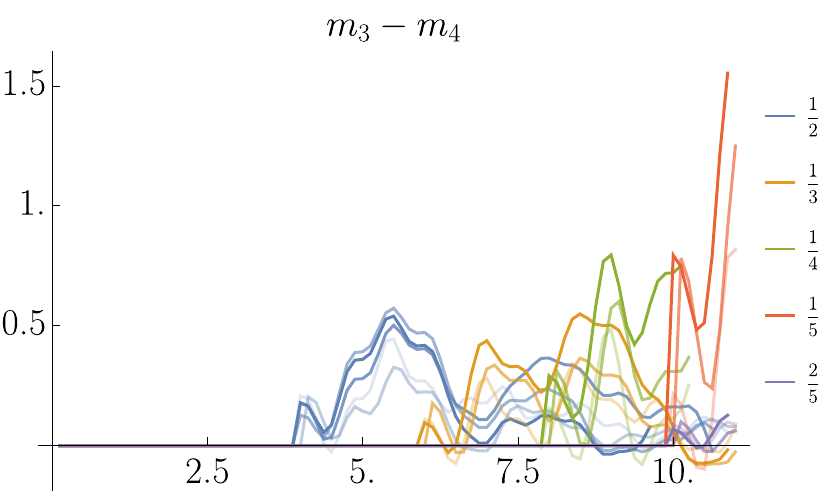}
    \end{minipage}
    \caption{Matching of the saddle multiplicities.
    The $x$-axis of all plots denotes the energy $s$. The $y$ axis is a bin average over explicit data of the amplitude, divided by the saddle prediction and tends to the corresponding saddle multiplicity. It tends to the multiplicity as indicated above the relevant plot. The first two plots are obtained from the decay width data, the second two from the full imaginary part while the last two are obtained from the full real part. The legends denote the value of $-\frac{t}{s}$. For the real part data, the weaker lines are the data computed with a smaller cutoff of $c_\text{max}$, which gives a sense of the convergence of the Rademacher procedure. There are sometimes some outliers in the plot such as for $-\frac{t}{s}=\frac{1}{2}$ in the second plot, since it can happen that the frequencies of the discrete Fourier transform accidentally coincide with other frequencies present in other saddles. Details about these plots can be found in the \texttt{Mathematica} notebook in our \texttt{Zenodo} submission \cite{Zenodo}.}
    \label{fig:Gross Mende multiplicity matching}
\end{figure}

\paragraph{Gross--Mende saddle.} We start with the largest one (which is the Gross--Mende saddle), determine its multiplicity and subtract it from the data and so on. In this way, we are able to constrain the multiplicities of the first 8 saddles with high confidence. Let us illustrate this procedure with the Gross--Mende saddle, for whose multiplicity we make the ansatz
\be 
\frac{m_1\, \mathrm{e}^{2\pi i s}+m_2 \, \mathrm{e}^{-2\pi i s}+m_3\, \mathrm{e}^{2\pi i t}+m_4 \mathrm{e}^{-2\pi i t}+m_3\, \mathrm{e}^{2\pi i u}+m_4 \mathrm{e}^{-2\pi i u}+m_5}{4 \sin^2(\pi s)} A_{\id}(s,\theta)\ .
\ee
We then determine these multiplicities by averaging these different data sets over bins to extract the relevant Fourier modes as described above. Not all datasets can resolve all the coefficients. From the decay widths we learn that $m_1+m_2+m_5=12$ and $m_3+m_4=0$, which are the first two figures in Figure~\ref{fig:Gross Mende multiplicity matching}. From the full imaginary part we also learn that $m_5=6$. The mass shifts tell us nothing new in this case because $A_{\id}(s,\theta)$ is purely imaginary. Finally, the full real part tells us that $m_1-m_2 \in \{4,6\}$\footnote{$m_1-m_2$ has to be even since $m_1+m_2$ is even.} and $m_3-m_4=0$. The data for the full real part decidedly worse than for the rest and these checks are by far the least conclusive. The plot to determine $m_1-m_2$ is particularly bad because it involves extracting the first Fourier mode of the Rademacher data. The error of the Rademacher data is small near integers which causes the error to oscillate with frequency 1, therefore contaminating the data. 

Luckily however, $m_2$ is actually determined by considering the analytic continuation in $s$ to $\Im s>0$, where the multiplcitiy has to approach $\pm 1$ (the sign depends on the choice of square root of the Hessian in the definition of $A_\id$). Only $m_2$ contributes to this limit and we learn that $m_2=\pm 1$. Together with the other constraints from the data, this fixes $m_1=5$ and $m_2=1$. This shows that $m_1-m_2=4$ is the correct value, which is matches also closest with the data in Figure~\ref{fig:Gross Mende multiplicity matching}. We should also note that convergence to the asymptotic behaviour is fastest for angles close to $\theta \sim \frac{\pi}{2}$, where the exponential fall-off is fastest. Therefore, only the curves with $-\frac{t}{s} \in \{\frac{1}{2},\frac{2}{5}\}$ match reasonably to this asymptotic in Figure~\ref{fig:Gross Mende multiplicity matching}.
\paragraph{Results.} We continue in the same way for the subleading saddles, for each saddle making the most general ansatz that is consistent with the growth condition. The data determines the multiplicities as follows. Unfixed multiplicities for which the data is not good enough to extract them are parametrized by constants $x_i^\gamma$. We always determine the contributions of $\gamma$, $C\gamma C$, $\gamma^{-1}$ and $C\gamma^{-1}C$ together since they are all of comparable magnitude. The result of this matching is given in Table~\ref{tab:numerical multiplicities}. Details can be found in the ancillary \texttt{Mathematica} notebook \verb|multiplicity_matching.nb| on \texttt{Zenodo} \cite{Zenodo}.
\begin{table}
    \centering
    \begin{tabular}{c|c|c}
    $\gamma$ & \text{elem.} & \text{multiplicities} \\
    \hline 
         $\begin{pmatrix}
        1 & 0 \\ 0 & 1
    \end{pmatrix}$ & \text{yes} & $(5\mathrm{e}^{2 \pi i  s}+\mathrm{e}^{-2 \pi i s} +6)A_\gamma$ \\
    \hline
    $\begin{pmatrix}
         1 & 0 \\ -2 & 1
     \end{pmatrix}$ & \text{yes} &
\rule{0pt}{6ex}$ \begin{aligned}&\mathrm{e}^{-\pi i  u} (\mathrm{e}^{-2 \pi i  s} x_1^{\gamma }+\mathrm{e}^{2 \pi i  s}+\mathrm{e}^{-2 \pi i  t} x_2^{\gamma }-\mathrm{e}^{2 \pi i u}
    x_2^{\gamma }+3-x_1^{\gamma})A_\gamma \\ &\quad+\mathrm{e}^{\pi i u} (\mathrm{e}^{2 \pi i s} (x_1^{\gamma }-3)+\mathrm{e}^{2\pi i t} x_2^{\gamma }-\mathrm{e}^{-2\pi i u}
   x_2^{\gamma }-x_1^{\gamma }+1)A_{\gamma^{-1}}\\
   &\quad+\text{cross}
    \end{aligned} $ \\
    \hline
    $\begin{pmatrix}
         1 & 0 \\ 4 & 1
     \end{pmatrix}$ & \text{no} & \rule{0pt}{6ex}$ \begin{aligned}&(\mathrm{e}^{2 \pi i  s} x_1^{\gamma }+\mathrm{e}^{-2 \pi i  s} x_2^{\gamma }-\mathrm{e}^{-2 \pi i t} x_3^{\gamma
   }+\mathrm{e}^{2 \pi i t} x_3^{\gamma }-\mathrm{e}^{-2 \pi i u} x_3^{\gamma }+\mathrm{e}^{2\pi i u} x_3^{\gamma }\\
   &\quad-x_1^{\gamma
   }-x_2^{\gamma })A_\gamma +(\mathrm{e}^{-2 \pi i s} x_1^{\gamma }+\mathrm{e}^{2 \pi i s} x_2^{\gamma }+\mathrm{e}^{-2 \pi i t} x_3^{\gamma
   }-\mathrm{e}^{2 \pi i t} x_3^{\gamma }\\
   &\quad+\mathrm{e}^{-2 \pi i u} x_3^{\gamma }-\mathrm{e}^{2 \pi i u} x_3^{\gamma }-x_1^{\gamma
   }-x_2^{\gamma})A_{\gamma^{-1}}+\text{cross} \\
   \end{aligned} $ \\
   \hline
   $\begin{pmatrix}
         1 & -2 \\ 0 & 1
     \end{pmatrix}$ & \text{yes} & \rule{0pt}{4.2ex} $\begin{aligned}&\mathrm{e}^{\pi i  s}  (\mathrm{e}^{-2 \pi i  s} (2-x_1^{\gamma })+2 \mathrm{e}^{2 \pi i s}+x_1^{\gamma }+4)A_{\gamma} \\
     &\quad+\mathrm{e}^{\pi is} (\mathrm{e}^{-2\pi i s} x_1^{\gamma }-x_1^{\gamma })A_{\gamma^{-1}}\end{aligned}$ \\
     \hline 
     $\begin{pmatrix}
         3 & 2 \\ 4 & 3
     \end{pmatrix}$ &\text{no} & $\mathrm{e}^{\pi i s} (\mathrm{e}^{-2 \pi i s}-1) x_1^{\gamma }A_\gamma-\mathrm{e}^{\pi i s} (\mathrm{e}^{-2 \pi i s}-1) x_1^{\gamma }A_{\gamma^{-1}}$ \\
     \hline
     $\begin{pmatrix}
         -7 & -4 \\ 2 & 1
     \end{pmatrix}$ & \text{yes} & \rule{0pt}{6ex}$ \begin{aligned}&\mathrm{e}^{ \pi i (s-t)} (-\mathrm{e}^{2\pi i  (t-s)} x_2^{\gamma }+\mathrm{e}^{-2 \pi i s} x_1^{\gamma }+\mathrm{e}^{2
   \pi i s}+\mathrm{e}^{2 \pi i t} x_2^{\gamma }+3-x_1^{\gamma }) A_\gamma \\
   &\qquad+\mathrm{e}^{-\pi i (s+t)}(\mathrm{e}^{2 \pi i (s+t)} (x_1^{\gamma }-x_2^{\gamma}-x_3^{\gamma }-1)+\mathrm{e}^{2\pi i
    s} x_3^{\gamma }\\
    &\qquad+\mathrm{e}^{2 \pi i t} (-x_1^{\gamma }+x_2^{\gamma }+x_3^{\gamma }+1)-x_3^{\gamma
   }) A_{\gamma^{-1}}+\text{cross} \end{aligned}$ \\
   \hline 
   $\begin{pmatrix}
         1 &0 \\ -6 & 1
     \end{pmatrix}$ & \text{no} & \rule{0pt}{8ex}$\begin{aligned}  &\mathrm{e}^{\pi i (s+t)}(\mathrm{e}^{-2 \pi i  (s+2 t)} x_2^{\gamma }+\mathrm{e}^{-2 \pi i (s+t)} x_4^{\gamma
   }+\mathrm{e}^{-2 \pi i s} x_1^{\gamma}-\mathrm{e}^{2 \pi i t} x_2^{\gamma }\\
   &\quad+\mathrm{e}^{-2 \pi i t} x_3^{\gamma }+x_5^{\gamma
   })A_\gamma+ \mathrm{e}^{\pi i (s+t)} (-\mathrm{e}^{-2 \pi i (s+2 t)} x_2^{\gamma }\\
   &\quad+\mathrm{e}^{-2 \pi i  (s+t)} (-x_1^{\gamma
   }-x_4^{\gamma }-x_6^{\gamma })+\mathrm{e}^{-2 \pi i s} x_6^{\gamma }+\mathrm{e}^{2\pi i t} x_2^{\gamma }\\
   &\quad+\mathrm{e}^{-2\pi i 
   t} (x_1^{\gamma }-x_3^{\gamma }+x_6^{\gamma })-x_1^{\gamma }-x_5^{\gamma }-x_6^{\gamma
   })A_{\gamma^{-1}}+\text{cross} \end{aligned}$ \\
   \hline 
   \rule{0pt}{6ex}$\begin{pmatrix}
         1 &0 \\ -8 & 1
     \end{pmatrix}$ & \text{no} &$\begin{aligned}&(-\mathrm{e}^{2\pi i (s+t)} x_3^{\gamma }+\mathrm{e}^{-2\pi i (s+t)} x_3^{\gamma }+\mathrm{e}^{2 \pi i s}
   x_1^{\gamma }+\mathrm{e}^{-2 \pi i s} x_2^{\gamma }-\mathrm{e}^{-2 \pi i  t} x_3^{\gamma }\\
   &\quad+\mathrm{e}^{2 \pi i t} x_3^{\gamma
   }+x_4^{\gamma }) A_\gamma +(\mathrm{e}^{2 \pi i (s+t)} x_3^{\gamma }-\mathrm{e}^{-2\pi i (s+t)} x_3^{\gamma }\mathrm{e}^{-2 \pi i s}
   x_5^{\gamma }\\
   &\quad+\mathrm{e}^{2\pi i s} x_6^{\gamma }+\mathrm{e}^{-2\pi i t} x_3^{\gamma }-\mathrm{e}^{2 \pi i t} x_3^{\gamma
   }+x_7^{\gamma })A_{\gamma^{-1}}+\text{cross}\end{aligned}$
    \end{tabular}
\caption{The numerically determined saddle multiplicities. We have stripped off the overall prefactor $(4 \sin^2(\pi s))^{-1}$ that is joint to all multiplicities. The saddles are roughly ordered according to their dominance. Uncertainties are parametrized by constants $x_i^\gamma$. The second column indicates whether this saddle is elementary. `cross' indicates the corresponding contribution where $t \leftrightarrow u$ and $\gamma \leftrightarrow C \gamma C$. For some of these saddles, accidents happen. E.g. for the fourth listed saddle satisfies $C \gamma C=U \gamma^{-1} U$, which implies that the contribution from $A_\gamma$ and $A_{\gamma^{-1}}$ are already crossing symmetric. Beyond the listed saddles, the numerical data becomes much worse and a reliable determination of the multiplicities is no longer possible.}
    \label{tab:numerical multiplicities}
\end{table}

 \paragraph{Conjectured multiplicities.} Let us consider \eqref{tab:numerical multiplicities} and make a few observations:
 \begin{enumerate}
     \item The data is consistent with only the elementary saddles contributing.
     \item The data is consistent with the multiplicities never involving exponentials other than $\mathrm{e}^{\pm 2\pi i s}$, apart from the overall prefactors.
     \item The data is consistent with all the principles that we mentioned above: multiplicities stabilize in the tail sum with a finite number of exceptions. In fact, this number of exceptions is very small as we shall discuss momentarily.
 \end{enumerate}
We will from on assume that these three principles indeed hold. This determines most of the unknowns. We can write the result at this point as
\begin{align}
    A&\sim \frac{1}{4 \sin^2(\pi s)}\Big[(5\mathrm{e}^{2 \pi i  s} +\mathrm{e}^{-2 \pi i s} +6)A_{\id}\nonumber\\
    &\qquad+ (\mathrm{e}^{-2 \pi i  s} x_1+\mathrm{e}^{2 \pi i  s}+3-x_1)(\mathrm{e}^{-\pi i  u}A_{\tilde{T}^{-2}}+\mathrm{e}^{-\pi i  t}A_{T^{-2}\tilde{T}^2}) \nonumber\\ 
    &\qquad+ (\mathrm{e}^{2 \pi i s} (x_1-3)-x_1+1)(\mathrm{e}^{\pi i u}A_{\tilde{T}^{2}}+\mathrm{e}^{\pi i t}A_{\tilde{T}^{-2}T^{2}}) \nonumber\\
    &\qquad+\mathrm{e}^{\pi i  s}  (\mathrm{e}^{-2 \pi i  s} (2-x_2)+2 \mathrm{e}^{2 \pi i s}+x_2+4)A_{T^{-2}} \nonumber\\
     &\qquad+\mathrm{e}^{\pi i s} (\mathrm{e}^{-2\pi i s}-1)x_2A_{T^2} \nonumber \\
     &\qquad+\mathrm{e}^{\pi i s} (\mathrm{e}^{-2 \pi i s} x_3+\mathrm{e}^{2
   \pi i s}+3-x_3) (\mathrm{e}^{-\pi i u}A_{\tilde{T}^{-2}T^{-2}}+\mathrm{e}^{-\pi i t}A_{T^{-4}\tilde{T}^2})\nonumber\\
   &\qquad+\mathrm{e}^{-\pi i s}(\mathrm{e}^{2\pi i
    s}-1) (x_3-1) (\mathrm{e}^{-\pi i u} A_{T^2 \tilde{T}^2}+\mathrm{e}^{-\pi i t} A_{\tilde{T}^{-2}T^4})+ \dots \Big]\ ,
\end{align}
where we renamed some of the unknowns. 

\begin{figure}
    \centering
    \includegraphics[width=0.5\linewidth]{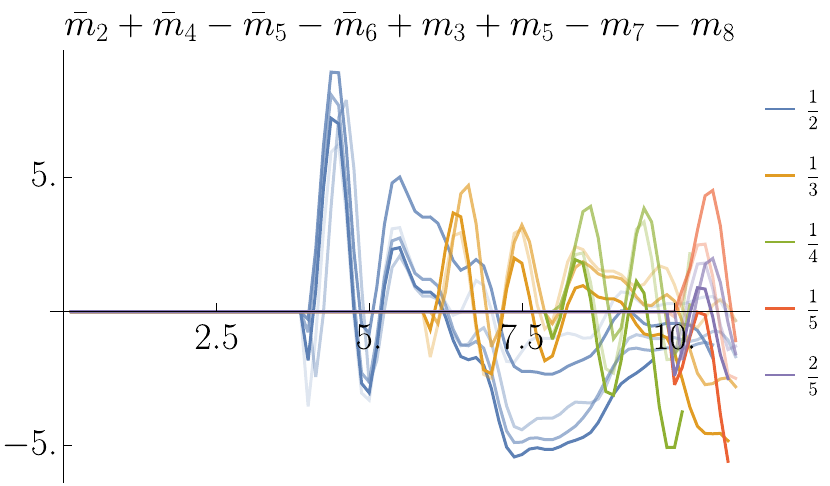}
    \caption{Extracting $\alpha$. This plot is obtained by extracting the corresponding Fourier mode of the full real data. This plot is similar to the last two plots in Figure~\ref{fig:Gross Mende multiplicity matching}. The limiting value is a certain combination of multiplicities in the ansatz. Once all the other determined multiplicities are inserted, the limit for large $s$ is expected to be $2(\alpha-1)$. Both limiting values $-2$ and $0$ seem consistent with the data, with a slight preference for $-2$, but the data is too imprecise, the Rademacher procedure has not converged enough and the energies are not high enough to claim this conclusively.}
    \label{fig:alpha plot}
\end{figure}

We recognize that this fits nicely the asymptotic form of \eqref{eq:tail sum over saddles}. Our analysis doesn't show how many exceptional terms one obtains in \eqref{eq:tail sum subtraction}. We can again make the most optimistic assumption that is consistent with the data, which suggests that $A_{T^2}$ can in fact be absent from the data (i.e.\ $x_2=0$) and $A_{T^2 \tilde{T}^2}$ can be absent (i.e.\ $x_3=1$). This then tells us that
\be 
M_\text{tail}(s,\theta)=\frac{\mathrm{e}^{2\pi i s}+\mathrm{e}^{-2\pi i s}+2}{4 \sin^2(\pi s)}=\cot^2(\pi s)\ .
\ee
After separating the tail contribution, we remain with the following final expression:
\begin{align}
    A &\sim \frac{1}{4\sin^2(\pi s)}\Big[(3\mathrm{e}^{2 \pi i  s}-\mathrm{e}^{-2 \pi i s} +2)A_{\id}+ (\mathrm{e}^{-2 \pi i  s}-1)\alpha (\mathrm{e}^{-\pi i  u}A_{\tilde{T}^{-2}}+\mathrm{e}^{-\pi i  t}A_{T^{-2}\tilde{T}^2}) \nonumber\\ 
    &\qquad+ (\mathrm{e}^{2 \pi i s} (\alpha-2)-\alpha)(\mathrm{e}^{\pi i u}A_{\tilde{T}^{2}}+\mathrm{e}^{\pi i t}A_{\tilde{T}^{-2}T^{2}})\nonumber\\
    &\qquad+(\mathrm{e}^{2\pi i s}+\mathrm{e}^{-2\pi i s}+2)\sum_{n=0}^\infty \mathrm{e}^{\pi i n s}\big(2A_{T^{-2n}} +\mathrm{e}^{-\pi i u}A_{\tilde{T}^{-2}T^{-2n}}+\mathrm{e}^{-\pi i t}A_{T^{-2n-2}\tilde{T}^{2}} \big)\Big]\ ,
    \label{eq:final formula}
\end{align}
where $\alpha=x_1-1$. We recall that $A_\gamma(s,\theta)$ was defined in \eqref{eq:Agamma expression}, where the action \eqref{eq:real part one loop action} and the determinant of the Hessian \eqref{eq:determinant Hessian} enters. The modular parameter $\tau_\theta$ is given through the kinematical data by \eqref{eq: tau theta definition}. The undetermined parameter $\alpha$ that we haven't been able to conclusively determine from the data doesn't show up at $s \in \ZZ$, nor does it show up in the imaginary part, nor is it fixed by general consistency conditions. The data seems to show a slight preference for $\alpha=0$, but this is not conclusive. The corresponding comparison with the data is given in Figure~\ref{fig:alpha plot}.

\section{Conclusion} \label{sec:conclusion}
In this paper, we revisited the saddle point evaluation of string amplitudes in the high-energy limit. We showed that it is necessary to  account for the contribution of complex saddles to correctly reproduce the asymptotic behavior. This leads to a far richer structure of amplitudes in the high-energy limit than what was previously assumed. In particular, we demonstrated that the additional saddles give rise to oscillatory contributions to the amplitude. At one-loop level, the possible phases are of the form $\mathrm{e}^{\pi i (m s+ n t)}$ with $m$ and $n \in \ZZ$.

Determining the multiplicities of the contributing saddles is a hard problem. We determined many of the multiplicities by comparing explicit numerical data with the saddle point expressions and invoking general constraints. This analysis was made possible because of the recent advancements in the calculation of the one-loop string amplitude through unitarity cut methods and the Rademacher evaluation introduced in \cite{Eberhardt:2022zay, Eberhardt:2023xck, Banerjee:2024ibt, Baccianti:2025whd}.  The result is summarized in Table~\ref{tab:numerical multiplicities}. Based on further favorable assumptions, we made a proposal for all multiplicities with the result  displayed in \eqref{eq:final formula}.

\paragraph{Higher genus.} We expect that many lessons of this paper extend to higher genera. In particular, it is natural to assume that the dominant saddles are given by surfaces $\mathcal{M}_{g,4}$ which are $g+1$-fold covers of the tree-level saddle, branched over four points where the vertex operators are inserted (even though we are not aware of a proof that there are no other candidate saddles). There exist analogous complex saddles also at higher genus, obtained by performing monodromies in the base surface for the holomorphic coordinates, as described in Section~\ref{subsec:complex Gross Mende saddles} for the one-loop case. Such complex saddles (or at least the equivalent of the elementary saddles) should contribute to higher genus amplitudes. Since the action is rescaled by a factor $(g+1)^{-1}$ with respect to the tree-level action, the phases of complex saddles are expect to be of the form $\mathrm{e}^{\frac{2\pi i}{g+1}(m s+n t)}$. Because the amplitude has a $(g+1)$-st order pole at $s \in \ZZ$, we now expect a common prefactor $(\sin(\pi s))^{-g-1}$ of the multiplicities. In the absence of reliable numerical computation schemes, we do not have a viable path towards determining the multiplicities of such saddle points at higher genus. We expect the presence of a similar tail sum as discussed in Section \ref{subsec:tail sum}, whose presence is dictated by the $i \varepsilon$ contour, together with approximate holomorphic factorization discussed in Section~\ref{subsec:elementary saddles}.

\paragraph{Implications for the non-perturbative amplitude.} It is interesting that the perturbative expansion of the full amplitude  necessarily breaks down at large energies. Indeed, the genus-$g$ amplitude scales like $g_\text{s}^{2g+2}\, \mathrm{e}^{-\frac{2s\mathcal{S}(\theta)}{g+1}}$, up to polynomial corrections. Thus, for large values of $s$, the string coupling is effectively necessarily strong. More precisely, for $g(g+1)\mathcal{S}(\theta)^{-1} \log (g_\text{s}^{-1}) \lesssim s \lesssim (g+1)(g+2)\mathcal{S}(\theta)^{-1} \log(g_\text{s}^{-1})$, the full non-perturbative amplitude is dominated by the genus-$g$ contribution. Therefore, for weak string coupling and $1 \ll s \lesssim 2\mathcal{S}(\theta)^{-1} \log(g_\text{s}^{-1})$, the non-perturbative amplitude is given to good approximation by the tree-level asymptotics \eqref{eq:tree-level asymptotics}. For $2\mathcal{S}(\theta)^{-1} \log(g_\text{s}^{-1})\lesssim s \lesssim 6\mathcal{S}(\theta)^{-1} \log(g_\text{s}^{-1})$, the non-perturbative amplitude is well-approximated by the main formula \eqref{eq:final formula} that we determined in this paper. Thus the asymptotics of the one-loop amplitude that we determine has a direct imprint on the full string amplitude.

\paragraph{Open strings.} It is an interesting problem to consider the behaviour of open strings at high energies. In \cite{Gross:1989ge}, the saddle point structure of open strings was discussed. These saddle points are the $\ZZ_2$-quotients of closed-string saddles and in particular the high energy limit of open string amplitudes is dominated by graviton exchanges. Thus, while details may differ, we expect that the physics that is probed by the analogous open string analysis is ultimately similar. On a computational level, it is actually simpler to carry out this analysis in the one-loop case, since the Rademacher formula is of lower complexity \cite{Eberhardt:2023xck}. At one-loop, the non-planar annulus diagram is expected to dominate over planar contributions since it features a closed string exchange. 

\paragraph{Boundedness in the complex plane.} We discussed a certain boundedness property of the amplitude in Section~\ref{subsec:analytic continuation}. We showed that the analytic continuation of the one-loop amplitude is bounded for $s$ in the first quadrant and away from the poles on the real axis while keeping the scattering angle $\theta \in(0,\pi)$ real. We also discussed the more involved statement that the same is true on the second sheet and also for $\Im s$ negative. Let us focus on the first property. Its proof was straightforward and applies at every genus, up to polynomial prefactors. In fact, the integrand at genus $g$ is polynomial of degree $4g-4$ in the momenta at genus $g$, since each picture-raising operator introduces one momentum.\footnote{$2n=8$ of these momenta get absorbed into the polarization structure that we factored out of the amplitude.} For large imaginary $s$, the integral can be done via saddle-point approximation and is dominated by saddles. The Hessian contributes a momentum dependence of the form $s^{3-3g-4}$. Thus, the amplitude decays like $s^{-g-3}$ along the imaginary direction in $s$. Together with the behaviour from saddle-point approximation, we can upgrade this with the help of the Phragm\'en-Lindel\"of principle. Thus we find that the amplitude obeys the following bound for $\Re s \ge 0$ and $\Im s \ge 1$:
\be 
\Big|A_g(s,\theta)s^{g+3}\mathrm{e}^{\frac{s\, \mathcal{S}(\theta)}{g+1}}\Big|\le  C_g(\theta)\ .
\ee
We unfortunately do not have good control of $C_g(\theta)$ (although it should presumably grow at least like $(2g)!$ with the genus), which prevents us from translating this into a non-perturbative bound.
Such a bound is fairly orthogonal to bounds that are usually applied in the S-matrix bootstrap, because it holds at fixed angle $\theta$. We hope that it may be useful to single out string theory in future bootstrap studies.

\paragraph{Chaos in the string amplitudes.} Chaotic behaviour of amplitudes of highly-excited strings has been a subject of recent interest \cite{Gross:2021gsj,Bianchi:2022mhs,Das:2023xge,Savic:2024ock}. Since loop amplitudes include such strings as intermediate states, it is interesting to ask whether our results suggest presence of chaos.

Our interpretation is that the high-energy limit of the one-loop string amplitude should not been seen as chaotic, since it exhibits regular oscillations on top of the exponential fall-off. A similar statement should be true for every fixed genus. However, once one resums the genus, all oscillation frequencies in are allowed, which will plausibly lead to a chaotic-looking answer. This argument is in agreement with ideas from black hole production. To observe black holes and their chaotic dynamics in the scattering process, one has to take large energies compared to the Planck length $\ell_\text{P}=g_\text{s}^{1/4} \ell_\text{s}$, not only with respect to the string length $\ell_\text{s}$ as we have done in this paper.

\section*{Acknowledgments}
We thank Nima Arkani-Hamed, Pinaki Banerjee, Maurizio Firrotta, Kelian H\"aring, Aaron Hillman, Juan Maldacena, Andrzej Pokraka for useful discussions.
MMB and LE are supported by the European Research Council (ERC) under the European Union’s Horizon 2020 research and innovation programme (grant agreement No 101115511). 
\appendix

\section{Gross--Mende toy integral}\label{app:toy integral}
In this appendix, we explain how one can efficiently compute the toy integral, to which the saddle point expression is compared to in Figure~\ref{fig: plots toy}. The toy integral is defined by \eqref{eq: definition of toy amplitude}.

\subsection{Exact evaluation of the imaginary part}\label{app: im part exact toy}
Set $\tau=x+i y$. The imaginary part \eqref{eq: toy integral im part formula} can be computed by taking the difference both $i \varepsilon$ prescriptions and reads
\begin{align}
    \Im \mathcal{A}^{\text{toy}} =  -\frac{1}{2i}\int_{-1}^{1}\d x \int_{y\in iL+\RR}\frac{\d y}{y^2} \, |\vartheta_2(x+iy)|^{-4s}|\vartheta_3(x+iy)|^{-4t}|\vartheta_4(x+iy)|^{-4u}  \ .
\end{align}
The strategy to evaluate this integral analytically is to employ the $q$-expansion of the integrand, similar to the corresponding computation for the full string amplitude \cite{Banerjee:2024ibt},
\be\label{eq: qexpansion integran toy}
\vartheta_2(\tau)^{-2s}\vartheta_3(\tau)^{-2t}\vartheta_4(\tau)^{-2u} = 2^{-2s}q^{-\frac{s}{4}}\sum_{n=0}^{\infty} a_n(s,\theta) q^{\frac{n}{2}} \ , \quad q=\mathrm{e}^{2\pi i \tau} \ .
\ee 
The $x$-integral will impose `level-matching', i.e.\ that $q$ and $\bar{q}$-coefficients in the expansion agree. It is then also simple to perform the $y$-integral, which leads to
\be 
   \Im \mathcal{A}^{\text{toy}} =2^{1-4s} \pi^2  \sum_{n=0}^{\infty} a_n(s,\theta)^2  \Theta(s-2n) (s-2n) \ , \label{eq:exact imaginary part toy integral}
\ee
where $\Theta(x)$ is the Heaviside step function.

\subsection{Numerical evaluation of the real part} \label{subapp:numerical evaluation}
The real part has to be evaluated numerically. A known strategy for this is described in \cite{Korpas:2019ava,Baccianti:2025gll}, where one performs the integral over a truncated fundamental domain where $y \le 1$ numerically and the integral over the remaining non-compact part of the fundamental domain analytically term-by-term in the $q$-expansion. However, there are exponentially large cancellations between both contributions for large values of $s$ which makes this integration procedure unsuitable. Instead, we perform the $y$ integral analytically, term-by-term in the $(q,\bar{q})$-expansion. Without loss of generality, we will focus only on evaluating the contribution of $\vartheta_2(\tau)^{-2s}\vartheta_3(\tau)^{-2t}\vartheta_4(\tau)^{-2u}$. We then add the 6 contributions \eqref{eq: definition of toy amplitude} in the end.

Using \eqref{eq: qexpansion integran toy}, the integral over $y$ of a given term in the $(q,\bar{q})$-expansion involves the integral
\begin{align}
    I(n,m,s)&:=\int_{\sqrt{1-x^2}}^\infty\d y \,\frac{q^{\frac{n}{2}-\frac{s}{4}} \bar{q}^{\frac{m}{2}-\frac{s}{4}} }{y^2} \\
    &=\mathrm{e}^{\pi i(n-m)\pi x}\bigg(\frac{\mathrm{e}^{-\pi(n+m-s)\sqrt{1-x^2}}}{\sqrt{1-x^2}}\nonumber\\
    &\hspace{3.5cm}-\pi(n+m-s)\Gamma\big(0,\pi(n+m-s)\sqrt{1-x^2}\big) \bigg) \ ,
\end{align}
where $\Gamma(a,b)$ is the incomplete Gamma function. We then perform the last integral numerically. By symmetry, it is enough to integrate over half of the fundamental domain,
\begin{multline}
    \int_{\mathcal{F}_{i\varepsilon}} \frac{\d^2 \tau}{\Im(\tau)^2} \,|\vartheta_2(\tau)|^{-4s}|\vartheta_3(\tau)|^{-4t}|\vartheta_4(\tau)|^{-4u} \\ = 2^{1-4s}\int_{0}^{\frac{1}{2}}\d x\sum_{m,n=0}^\infty a_n(s,\theta)a_m(s,\theta) I(n,m,s) \ .
\end{multline}
We checked that the imaginary part of the result agrees numerically with the exact expression \eqref{eq:exact imaginary part toy integral}.

\bibliographystyle{JHEP}
\bibliography{bib}
\end{document}